\documentclass[aps,prl,twocolumn,superscriptaddress,floatfix]{revtex4-1}

\usepackage{graphicx}
\usepackage{blindtext}
\usepackage{floatrow}
\usepackage{lipsum}
\usepackage{color}
\usepackage[dvipsnames]{xcolor}
\usepackage{amsmath}
\usepackage{amssymb}
\usepackage{wasysym}
\usepackage{epstopdf}
\usepackage{fdsymbol}
\usepackage{pdfpages}
\usepackage{appendix}
\usepackage{pgffor}
\usepackage{physics}
\usepackage{bm}
\usepackage{ulem}

\newif\ifarXiv
\arXivtrue 

\makeatletter
\AtBeginDocument{\let\LS@rot\@undefined}
\makeatother

\begin{document}

\preprint{APS/123-QED}

\title{Active Microrheology of Colloidal Suspensions of Hard Cuboids}

\author{Effran Mirzad Rafael}
\affiliation{Department of Chemical Engineering, The University of Manchester, Manchester, M13 9PL, United Kingdom}%
\author{Luca Tonti}
\affiliation{Department of Chemical Engineering, The University of Manchester, Manchester, M13 9PL, United Kingdom}%
\author{Fabi\'an A. Garc\'ia Daza}
\affiliation{Department of Chemical Engineering, The University of Manchester, Manchester, M13 9PL, United Kingdom}%
\author{Alessandro Patti}
\email{a.patti@ugr.es}
\affiliation{Department of Chemical Engineering, The University of Manchester, Manchester, M13 9PL, United Kingdom}
\affiliation{Department of Applied Physics, University of Granada, Avenida Fuente Nueva s/n, 18071 Granada, Spain}%


\begin{abstract}
By performing dynamic Monte Carlo simulations, we investigate the microrheology of isotropic suspensions of hard-core colloidal cuboids. In particular, we infer the local viscoelastic behaviour of these fluids by studying the dynamics of a probe spherical particle that is incorporated in the host phase and is dragged by an external force. This technique, known as active microrheology, allows one to characterise the microscopic response of soft materials upon application of a constant force, whose intensity spans here three orders of magnitude. By tuning the geometry of cuboids from oblate to prolate as well as the system density, we observe different responses that are quantified by measuring the effective friction perceived by the probe particle. The resulting friction coefficient exhibits a linear regime at forces that are much weaker and larger than the thermal forces, whereas a non-linear, force-thinning regime is observed at intermediate force intensities.

\end{abstract}


\maketitle


\section{Introduction}
Complex fluids are present in a wide variety of day-to-day consumables, including cosmetics, pharmaceuticals, paints and foods. A common thread shared by these families of products is the correlation between the interactions established at the sub-micron scale and their macroscopic response as well as the ability of their microscopic domains to rearrange upon the application of external stimuli \cite{cicuta2007microrheology}. When flow is imposed, complex fluids tend to display behaviours characteristic of non-Newtonian fluids, such as viscoelasticity \cite{zia2018active}. Traditionally, the rheology of fluids has been studied with mechanical rheometers, providing information mainly on their bulk flow behaviour. However, advances in rheological techniques have enabled the assessment of the viscoelastic behaviour of soft materials through a technique referred to as microrheology (MR) \cite{cicuta2007microrheology}. MR operates under the mechanism of embedding a colloidal tracer in a host fluid, whose flow properties can be inferred by analysing the tracer's dynamics. This technique has been applied to study a wide spectrum of systems including chromonic liquid crystals \cite{habibi2019passive}, DNA \cite{chapman2014nonlinear, fernandez2018microrheology}, actin networks \cite{levin2020kinetics}, biofluids \cite{weigand2017active,watts2013investigating}, hard \cite{wilson2009passive} and soft \cite{Nazockdast2016} spheres. More specifically, passive MR involves monitoring the response of the tracer merely due to the thermal fluctuations of the host fluid (or bath) - this will probe the linear response of the complex fluid. By contrast, active MR unveils the fluid's nonlinear response by applying an external force to the tracer - active MR can be performed at fixed force or fixed velocity. In fixed-forced active MR, the tracer is pulled with a constant force through the bath and, by applying the Stoke's drag law, the effective friction coefficient (or microviscosity) of the bath can then be obtained \cite{carpen2005microrheology}. 

Among complex fluids, colloidal suspensions of anisotropic particles are especially attractive for their rich phase behaviour and ability to self-assemble in a wide spectrum of mesophases, such as nematic and smectic liquid crystals (LCs). In particular, biaxial particles, such as boards, ellipsoids, rhombuses and bent-core particles, are able to form exotic LC phases \cite{dussi2018hard, tasios2017simulation, chiappini2019biaxial, fernandez2020shaping, berardi2000thermotropic}, including the biaxial nematic ($\rm N_{B}$) phase, whose existence at molecular scale is still object of discussions \cite{lehmann2019molecular}. As far as nanoboards are concerned, the first experimental evidence of the existence of stable $\rm N_{B}$ phases was reported about a decade ago in dispersions of mineral goethite particles \cite{van2009experimental}. This discovery has inspired further works on the phase behaviour of board-like particles, which has been extensively studied by experiments, theory and simulations \cite{van2010onsager, van2010isotropic, martinez2011biaxial, belli2011polydispersity, belli2012depletion, peroukidis2013phase, peroukidis2013supramolecular, peroukidis2014biaxial, mederos2014hard, gonzalez2015effect, cuetos2017phase, yang2018synthesis, patti2018monte, cuetos2019biaxial, rafael2020self, skutnik2020biaxial}. Nevertheless, less attention has been given to the study of their dynamics, so far mostly limited to the long-time structural relaxation of uniaxial nematic LCs \cite{cuetos2020dynamics},  field-driven uniaxial-to-biaxial nematic switching \cite{mirzad2021dynamics} and in cylindrical confinement \cite{patti2021dynamics}. Generally speaking, it was found that the dynamics of board-like particles strongly depends on their geometry, which spans prolate (rod-like) to oblate (disk-like) shapes, and interesting behaviours are observed at the self-dual shape, where oblate and prolate geometries fuse into one. In particular, the self-dual nanoboards were found to exhibit the lowest overall translational diffusivity in equilibrium uniaxial nematic phases; and show the slowest overall response time in field-induced uniaxial-biaxial nematic switching due to biaxial retention tendencies, when compared to other geometries. These findings highlight the importance of shape anisotropy to control the dynamics of board-like particles. Currently, state-of-the-art commercial displays are engineered with molecular LCs, and colloidal LCs serve mainly as model systems to understand the behaviour of their molecular counterparts. However, colloidal LCs are regarded as potential candidates for next-generation displays, as these systems are athermal, relatively cheap and highly susceptible to external fields \cite{gabriel2000, tschierske2010biaxial, Lekkerkerker2013, Leferink2014}. As such, understanding the dynamics and rheology of these systems is as relevant as mapping their phase behaviour.

To this end, in this work, we employ the dynamic Monte Carlo (DMC) simulation technique to investigate the viscoelastic response of suspensions of hard cuboids by active MR. The DMC method has been recently shown to reproduce MR results of Langevin Dynamics in systems of spherical and rod-like particles to an excellent degree of qualitative and quantitative agreement \cite{daza2021microrheology}. In particular, we are interested in the nonlinear response of board-like particles, as application and manufacturing processes require complex fluids to flow and be driven out-of-equilibrium. More specifically, we investigate the effective friction of a bath of board-like particles in the isotropic phase through fixed-force active MR. The investigation will be framed within the effects of altering the packing fraction of the bath and the particle geometry (board-like particles are morphed from prolate to oblate). The present paper is arranged as follows: we first discuss the details of our model systems and simulation methods, referring the reader to our recent work for details on how DMC has been adapted to study active MR \cite{daza2021microrheology}, then we analyse the results at different P\'eclet numbers when varying either the system packing or the geometry of the bath particles, and, in the final section, we draw our conclusions.


\section{Model and Simulation Methodology}

\begin{figure}[h!]
\centering
\includegraphics[width=\columnwidth]{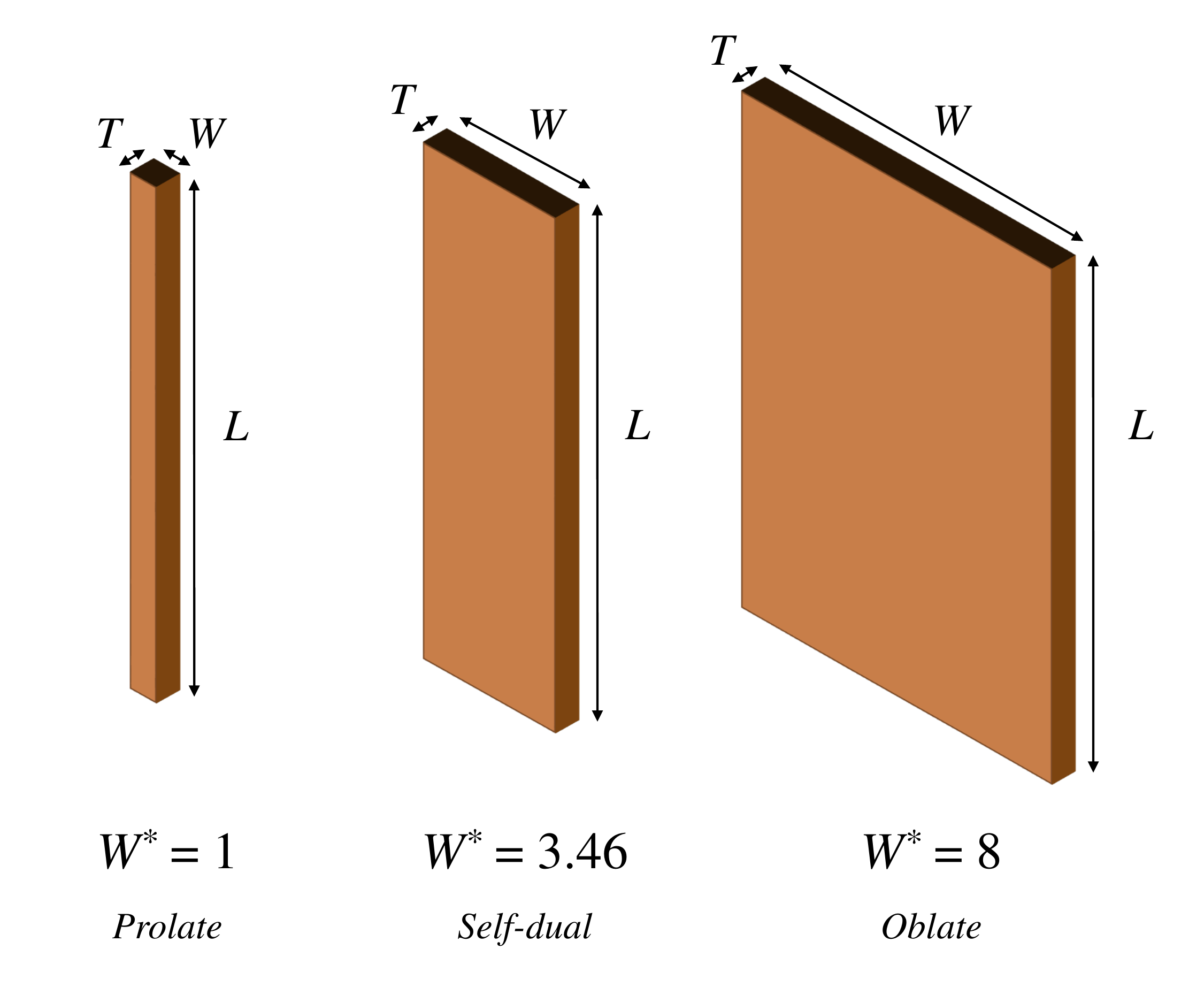}
\caption{Model HBPs studied in this work. Length, width and thickness are respectively labelled as $L=12T$, $W$ and $T$, with $T$ the system unit length. The reduced width, $W^* \equiv W/T$, assumes three different values to reproduce prolate, self-dual and oblate geometries.}
\label{cuboids}
\end{figure}

Our systems consist of $N_{c} = 1000$ hard board-like particles (HBP) and $N_{s}$ = 1 hard spherical tracer, constrained in an elongated box with volume $V_{box} = L_{x} \times L_{y} \times L_{z}$, where $L_{x} = L_{y}$ and $L_{z} = 3L_{x}$. HBPs are cuboids of thickness $T$, length $L$ and width $W$, with $T$ the system unit length. In this work, all HBPs have a reduced length given by $L^* \equiv L/T = 12$, while the reduced width, $W^* \equiv W/T$, is a simulation parameter that takes the values $W^*=\{1, \sqrt{L^*} \approx 3.46, 8\}$, providing respectively prolate, self-dual shaped and oblate geometries. The spherical tracer has a diameter $\sigma=T$. A schematic representation of these particles is given in Fig.\,\ref{cuboids}. Given the hard-core nature of the particles, the phase behaviour of the systems is fully characterised by the particle geometry and system packing fraction, which can be approximated to

\begin{equation} \label{equation:eta}
    \phi \approx \frac{N_{c}v_{0}}{V}
\end{equation}

\noindent where $v_{0}=TWL$ is the volume of one HBP and the contribution of the spherical tracer has been disregarded \cite{tonti2021diffusion}. In the first part of this work, we will assess the impact of $\phi$ on the effective friction of the bath. In this part, we employ isotropic (I) phases consisting of HBPs at a fixed width, $W^{*} = 3.46$, and comparisons are made across three packing fractions, namely $\phi$ = 0.20, 0.25 and 0.30. The reason why we selected the self-dual shape to ponder the effect of density is due to the relatively large stability range of the I phase at this specific particle geometry \cite{cuetos2017phase}. In the second part, we are interested in altering the HBP's geometry from prolate to oblate at a fixed packing fraction. In this case, we will keep the packing fraction constant at $\phi = 0.20$ and increase the particle width from $W^{*} = 1$ to 8.

To equilibrate the systems we employed standard MC simulations in the canonical ensemble, reproducing the I phases from our previous work \cite{cuetos2017phase}. Because the interactions between particles are modelled by a hard-core potential, moves are always accepted unless an overlap is produced. Overlaps between cuboids are checked with the separating axes theorem proposed by Gottschalk and coworkers \cite{gottschalk1996obbtree} and later adapted by John and Escobedo to study the phase behaviour of HBPs with square cross section \cite{john2005phase}. To check the occurrence of overlaps between the spherical tracer and cubes, we employed the OCSI algorithm \cite{tonti2021fast}. Equilibration was considered achieved when the uniaxial order parameters showed a steady value within moderate statistical fluctuations. Because we are only investigating MR in I phases, equilibration was relatively fast, usually taking no more than $1 \times 10^{6}$ MC cycles, with each cycle consisting of $N=N_c+N_s$ attempts of displacing or rotating a randomly selected particle. In particular, the uniaxial order parameters have been obtained from the diagonalisation of the following symmetric tensor:

\begin{equation}
\textbf{Q}^{\lambda \lambda} = \frac{1}{2N} \Biggl\langle \sum_{i=1}^{N} \big(3\hat{\lambda}_{i} \cdot \hat{\lambda}_{i} - \textbf{I}\big) \Biggr\rangle
\end{equation}

\noindent where $\hat{\lambda}_i$ = $\hat{\textbf{x}}$, $\hat{\textbf{y}}$, $\hat{\textbf{z}}$ are unit vectors respectively aligned with \textit{W}, \textit{T} and \textit{L}, while \textbf{I} is the identity tensor. The diagonalisation of $\textbf{Q}^{\lambda \lambda}$ results in three eigenvalues ($S_{2,W}$, $S_{2,T}$, $S_{2,L}$) and their corresponding eigenvectors ($\rm \hat{\textbf{m}}$, $\rm \hat{\textbf{l}}$, $\rm \hat{\textbf{n}}$). A phase is considered nematic if at least one of the eigenvalues is larger than 0.40. In our equilibrated phases, all the uniaxial order parameters are below the threshold required to be classified as nematics. The resulting $\rm I$ phases are then used as initial configurations in the DMC production runs.

The DMC method produces Brownian Dynamics (BD) trajectories by rescaling an arbitrarily set MC time step, $\delta t_{\rm MC,m}$, with the acceptance rate, $\mathcal{A}_m$, where $m = c,s$ refers to the cuboidal bath particles and spherical tracer, respectively. The interested reader is referred to our past works for further details on the DMC technique \cite{patti2012brownian, cuetos2015equivalence, corbett2018dynamic, daza2020dynamic, chiappini2020, daza2021microrheology}. Here, we only review the most relevant aspects that are instrumental to the present work. Essentially, we set $\delta t_{\rm MC,s}$ within values between $10^{-2}\tau$ and $10^{-7}\tau$, where $\tau = \eta T^{3}/(k_{B}T_{b})$ is the time unit, while $T_b$ and $\eta$ are, respectively, the bath temperature and viscosity. As a result, the MC time step of the tracer is obtained through the corresponding acceptance rates (see Supplemental Material at [URL] for details). More specifically, when the tracer is subjected to a one-dimensional external force in the positive direction of $\hat{\textbf{z}}$, taking the form $\textbf{F}_{\rm ext} = F_{\rm ext} \hat{\textbf{z}}$, the time step $\rm \delta t_{MC,s}$ is obtained from the following relationship \cite{daza2020dynamic}:

\begin{equation}
    \bigg( \frac{3}{2}\mathcal{A}_{s} - \frac{1}{2}\bigg)\;\delta t_{\rm MC,s} = \mathcal{A}_{c} \delta t_{\rm MC,c}
\end{equation}

\noindent where $\mathcal{A}_{s}$ and $ \mathcal{A}_{c}$ are the acceptance rates of spherical tracer and cubes, respectively. It should be noted that the external force is considered in the above equation by means of $\mathcal{A}_s$. According to our previous work, $\mathcal{A}_s\sim 1-\beta F_{ext}\delta x_z/4$ \cite{daza2020dynamic}. In active MR-DMC simulations, we stress that in order to produce the most reliable approximations, the following condition should be satisfied:

\begin{equation}
  \beta F_{\rm ext} \delta x_{z} \ll 1  
\end{equation}

\noindent where $\beta=1/k_B T_b$, with $k_B$ the Boltzmann’s constant and $\delta x_{z}$ the maximum displacement of the tracer in the direction of the force. The so-equilibrated $\delta t_{\rm MC,s}$, along with $\delta t_{\rm MC,c}$, are then used to produce the time trajectories by DMC simulations. We applied periodic boundary conditions to our elongated simulation boxes and do not perform unphysical moves such as jumps, swaps and cluster moves in order to produce the correct dynamics. The tracer is pulled by an external force $\textbf{F}_{\rm ext}$ parallel to the z-axis that takes the form:

\begin{equation}
    \beta \textbf{F}_{\rm ext} =  \frac{Pe}{a} \hat{\textbf{z}}
\end{equation}

\noindent where $a = \sigma / 2$ is the tracer radius, and $Pe$ is the P\'eclet number which gives the ratio of advection to thermal forces. The displacement of the HBPs' centres of mass, $\delta \textbf{r}_{i}$, is decoupled into three terms: $\delta \textbf{r}_{c,i} = X_{W}\hat{\textbf{u}} + X_{T}\hat{\textbf{v}} + X_{L}\hat{\textbf{w}}$, where $X_{\alpha}$, with $\alpha = W, T, L$, is the maximum displacement allowed to a generic HBP. These displacements are set by the following Einstein relations:

\begin{equation}
    \abs{X_{\alpha}} \leq \sqrt{2D^{\rm tra}_{\alpha,i}\delta t_{\rm MC,c}},
\end{equation}

\noindent where $D^{\rm tra}_{\alpha,i}$ are the translational diffusion coefficients at infinite dilution along the three particle direction. Similarly, the rotation of the HBPs are performed \textit{via} three consecutive rotations with maximum rotation $Y_{W}$, $Y_{T}$ and $Y_{L}$ around $\hat{\textbf{u}}$, $\hat{\textbf{v}}$ and $\hat{\textbf{w}}$, respectively. These maximum rotations read

\begin{equation}
    \abs{Y_{\alpha}} \leq \sqrt{2D^{\rm rot}_{\alpha,i}\delta t_{\rm MC,c}},
\end{equation}

\noindent where $D^{\rm rot}_{\alpha,i}$ are the rotational diffusion coefficients at infinite dilution. Both translational and rotational diffusion coefficients have been calculated with the open-source software HYDRO++ \cite{carrasco1999hydrodynamic, garcia2007improved}. For the specific values of $D^{\rm tra}_{\alpha,i}$ and $D^{\rm rot}_{\alpha,i}$, readers are referred to our previous work on the equilibrium dynamics of HBPs \cite{cuetos2020dynamics}, where values are reported in units of $D_{0} \equiv T^{2}\tau^{-1}$ (translational) and $D_{r} \equiv$ rad$^{2}\tau^{-1}$ (rotational).

As far as the spherical tracer is concerned, we disregard rotations and only consider translational moves. In particular, the displacement of the tracer's centre of mass reads $\delta \textbf{r}_{s} = Z_{\parallel}\hat{\textbf{i}} + Z_{\perp}\hat{\textbf{j}} + Z_{\perp}\hat{\textbf{k}}$, where $\hat{\textbf{i}}$ is the displacement vector parallel to the external force $\textbf{F}_{\rm ext}$ while $\hat{\textbf{j}}$ and $\hat{\textbf{k}}$ are vectors orthogonal to $\hat{\textbf{i}}$ and to each other. Due to the presence of the external force, the resulting maximum displacement of the spherical tracers incorporates two contributions and reads

\begin{equation} \label{zdirection}
    | Z_{\parallel} | \leq \sqrt{2D_{s} \delta t_{\rm MC,s} + (D_{s} \beta F_{\rm ext} \delta t_{\rm MC,s})^2}
\end{equation}

\noindent while the maximum displacement in planes perpendicular to $\textbf{F}_{\rm ext}$ is similar to that of HBPs:

\begin{equation}
    | Z_{\perp} | \leq \sqrt{2D_{s} \delta t_{\rm MC,s}}
\end{equation}

\noindent The tracer diffusion coefficient at infinite dilution, $D_s$, is obtained from the Stokes-Einstein relation:

\begin{equation}
    \frac{D_{s}}{D_{0}} = \frac{1}{3\pi}
\end{equation}

\noindent In this paper, we are interested in the rheology of HBPs in the I phase. To evaluate this, we computed the effective friction coefficient derived from the Stokes-drag expression:

\begin{equation}
    \frac{\gamma_{\rm eff}}{\gamma_{0}} = \frac{F_{\rm ext}}{6 \pi \eta a \langle v_{s} \rangle}
\end{equation}

\noindent where $\gamma_{0} = 6 \pi \eta a$ is the friction coefficient of the medium and $\langle v_{s} \rangle$ is the mean velocity of the tracer at long times. The initial $\rm I$ phases were equilibrated in a way to ensure that all configurations were mutually uncorrelated. Following equilibration, in the DMC production run, we applied an external force of magnitude $(0,0,F_{\rm ext}) = (0,0,Pe\:k_B T_b/a)$ to the tracer and allowed it to displace a distance of at least $L_{z}/2$ (half of the longest box length). We set between $2 \times 10^{5}$ to $4 \times 10^{6}$ MC cycles for our simulations, depending on the values of $\phi$ and $Pe$. All data points are averaged from 750 independent trajectories.


\section{Results}

\begin{figure*}[htbp!]
\centering
    \includegraphics[width=0.99\linewidth,height=0.26\textheight]{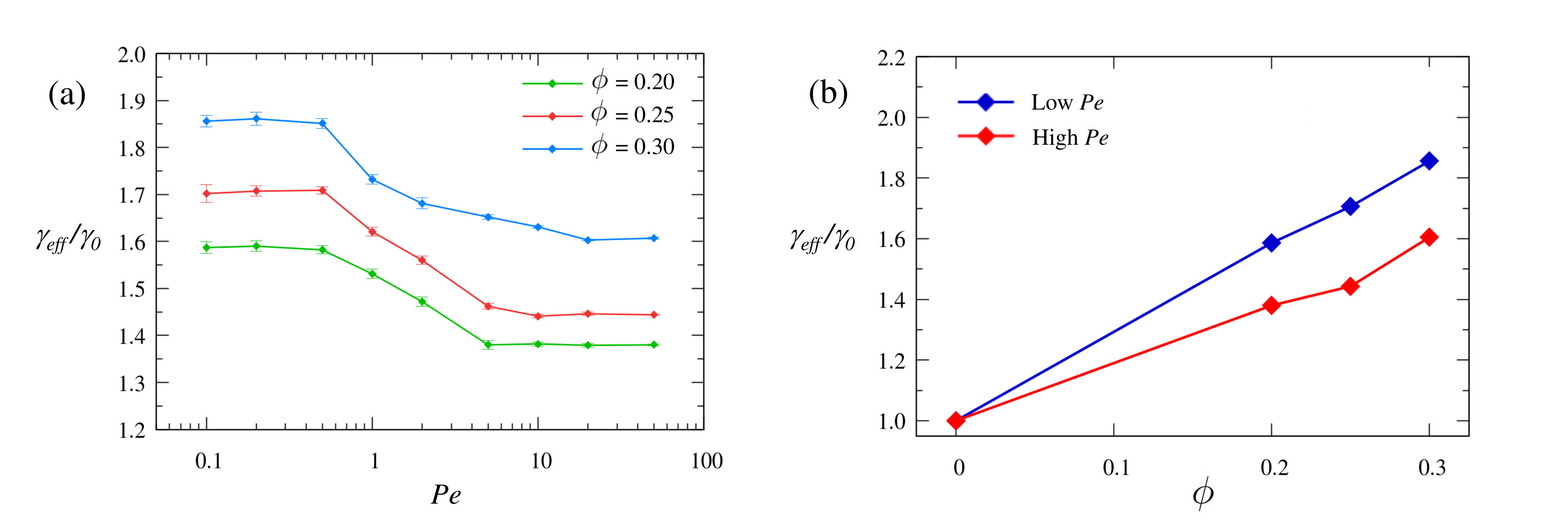}
   \caption{(a) Dependence of the friction coefficient with $Pe$ in I phases of HBPs with $L^{*} = 12$ and $W^{*} = 3.46$ at $\phi = 0.20$, 0.25 and 0.30. (b) Variation of $\gamma_{\rm eff}/\gamma_{0}$ at high and low $Pe$ at different $\phi$. The lines are guides for the eye.}
    \label{Ch6_Fig1}
\end{figure*}

We first report the effect of increasing the system packing fraction on the effective friction coefficient in I phases of self-dual-shaped HBPs. At this particular geometry, where $W=\sqrt{LT} \approx 3.46T$, the I phase is stable up to $\phi=0.30$. At any other particle width, within the range $1 \le W^* \le 12$ and $\phi=0.30$, oblate or prolate nematic LCs are observed \cite{cuetos2017phase}. The dependence of the friction coefficient on the $Pe$ number for this family of HBPs is reported in Fig.~\ref{Ch6_Fig1}(a) for $\phi = 0.20$, 0.25 and 0.30. While the friction experienced by the probe particle increases with the system density, its qualitative behaviour is very similar and characterised by the presence of three separate regimes.

\begin{figure*}[htbp!]
\centering
    \includegraphics[width=0.75\linewidth,height=0.61\textheight]{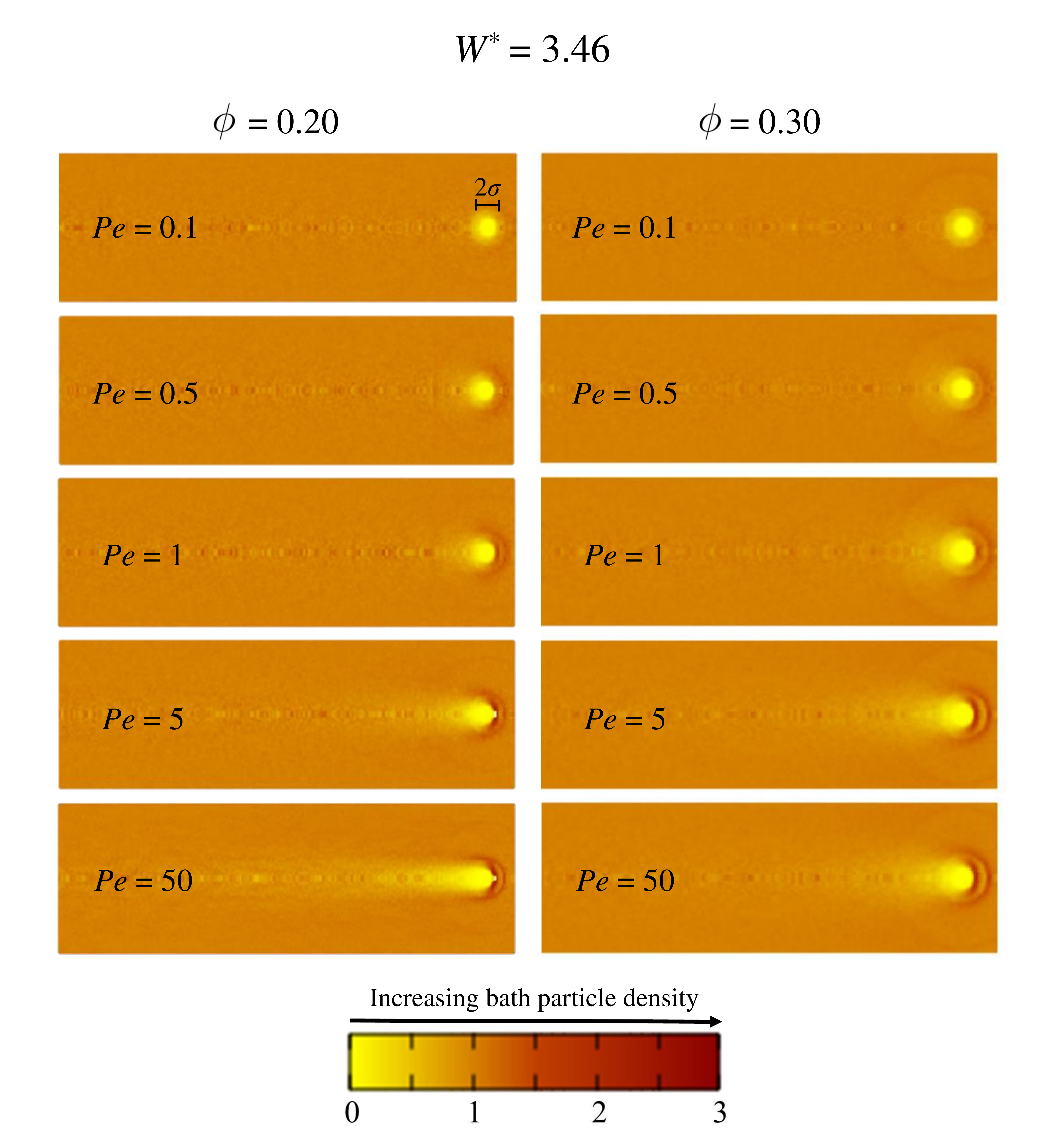}
   \caption{Density maps of self-dual shaped HBPs at $\phi = 0.20$ (left) and $\phi = 0.30$ (right) and at the Pe numbers indicated in each frame. The colour palette is shown on at the bottom of the figure and refers to the ratio between the local density and the bath density. Yellow regions indicate low bath particle density, while the dark red regions indicate high bath particle density.}
    \label{Ch6_Fig3}
\end{figure*}

At $Pe<0.5$, corresponding to relatively weak external forces, a plateau is observed, with $\rm \gamma_{eff}/\gamma_0 \approx 1.6$, 1.7 at 1.85 and $\phi=0.20$, 0.25, and 0.30, respectively. At such small $Pe$ numbers, the advection force exerted by the tracer is too weak to significantly perturb the local particle distribution of the host HBPs, merely leading to weakly distorted microstructures in the bath \cite{squires2005simple}. Basically, the thermal fluctuations of the surrounding fluid dominate on the external force applied, leading to an essentially symmetric distribution of HBPs around the tracer. This can be clearly observed by calculating the local changes in density $\rho$ of bath particles within volumes $v$ around the tracer. Since the external force induces an axially symmetrical distribution of host particles near the tracer, the volumes $v$ are defined by virtually dividing the space into an arbitrary set of concentric rings centered on the tracer axis. We have calculated the local densities as $\rho(v)=\langle N(t)/v\rangle$  where $N(t)$ refers to the number of particles in volume $v$ at time $t$, and $\langle\ldots\rangle$ indicates time average. As shown in Fig.\,\ref{Ch6_Fig3}, at $Pe = 0.1$ the density distribution of HBPs around the tracer is  uniform at both $\phi = 0.20$ (left column) and $\phi = 0.30$ (right column). When the advection force is increased slightly to $Pe = 0.5$, the density of HBPs surrounding the tracer is still relatively uniform, and we expect the effective friction coefficient to be similar to $Pe = 0.1$. This low-$Pe$ linear regime approaches the passive microrheology limit \cite{khair2006single, squires2005simple}. We also note that the presence of the probe particle does not have a tangible effect on the orientation of the cuboids around it, which maintain their random orientation, similarly to the rest of the bath particles. This does not exclude that some of them change orientation while the tracer is close enough, but the global effect does not indicate substantial alignment. We have recently investigated this scenario in I phases of hard spherocylinders by calculating a local orientational correlation function and found a very weak, negligible ordering at low Pe numbers, which fades out as soon as the applied force increases and the non-linear regime is approached \cite{daza2021microrheology}. We indeed see an increase in density around the tracer as revealed by the density maps shown in Fig.\,\ref{Ch6_Fig3}, that agree well with our findings in the I phase of hard spherocylinders. However, this increase in local density is not accompanied by an increase in ordering: the cuboids around the tracer basically remain randomly oriented.

At $Pe \geq 1$, advection forces begin to dominate over thermal forces and a force-thinning regime, with the friction coefficient decreasing with increasing the intensity of the force applied, develops. This non-linear regime closely reminds the shear-thinning observed in non-Newtonian fluids, whose viscosity decreases as the shear rate increases. It spans over $1 \leq Pe \leq 5$ for $\phi = 0.20$ and $0.25$, and over $1 \leq Pe \leq 20$ for $\phi = 0.30$. The occurrence of a force-thinning regime is in line with what has been reported by theoretical predictions \cite{squires2005simple, khair2006single, gazuz2009active, swan2013active}, simulations \cite{puertas2014microrheology, carpen2005microrheology, daza2020dynamic} and experiments \cite{meyer2006laser, sriram2010active}. In this regime, the tracer has sufficient driving force to induce stronger microstructural distortions and cause a symmetry breaking of the HBPs surrounding the tracer. In the density maps shown in Fig.\,\ref{Ch6_Fig3}, at both $\phi = 0.20$ and 0.30, we can see the onset of accumulation of bath particles, represented by a dark blue cap in the upstream face of the tracer. The same density maps reveal the existence of a low-density trail (or wake) that forms behind the tracer, indicating that HBPs need some time to heal the distortions caused by the forced displacement of the tracer. In agreement with past theoretical, simulation and experimental works, this depletion trail increases in length as $Pe$ increases \cite{swan2013active, sriram2010active, carpen2005microrheology}. In this regime, the tracer's mobility increases as the reduction in the effective friction coefficient suggests. It is also noted that the force-thinning regime in I phases of self-dual HBPs spans a much smaller range, especially when compared to systems of (quasi-)hard spheres \cite{puertas2014microrheology, carpen2005microrheology} or hard spherocylinders \cite{daza2021microrheology}. For instance, at $\phi = 0.30$, the force-thinning regime for hard spherocylinders spans $2 \leq Pe \leq 50$ while for (quasi-)hard spheres, it is $2 \leq Pe \leq 100$. This trend seems to indicate that as the host particles become more anisotropic, the force-thinning regime tends to span a smaller range with the effective friction coefficients also generally becoming smaller. This can be expected as host particles that are more anisotropic are larger and harder to distort; therefore, the tracer's mobility becomes more hindered. 

Finally, at larger $Pe$ numbers, the effective friction coefficient exhibits a second plateau. This plateau corresponds to the high-$Pe$ regime, where the advection force dominates thermal forces \cite{khair2006single,squires2005simple}. When analysing the density distribution in Fig.\,\ref{Ch6_Fig3}, at $Pe = 50$ (bottom row), the trail of depleted particles is longer for a system that is less packed. More specifically, at $\phi = 0.20$, the trail's length is approximately $5\sigma$, whereas at $\phi = 0.30$, it is about $3\sigma$. This phenomenon is most likely due to the concentration gradient of the bath that forms around the tracer. At $\phi = 0.30$, this concentration gradient is larger than that at $\phi = 0.20$ and hence drives particles back into the depleted trail faster than the concentration gradient forming at $\phi = 0.20.$ \cite{gazuz2009active}. We stress, however, that we have neglected hydrodynamic interactions (HI). The theoretical formalism developed by Khair and Brady suggested that if HI effects are significant, force-thickening may occur at high $Pe$, and the effective friction coefficient may experience an increase \cite{khair2006single, zia2018active}.

\begin{figure*}[htbp!]
\centering
    \includegraphics[width=0.99\linewidth,height=0.26\textheight]{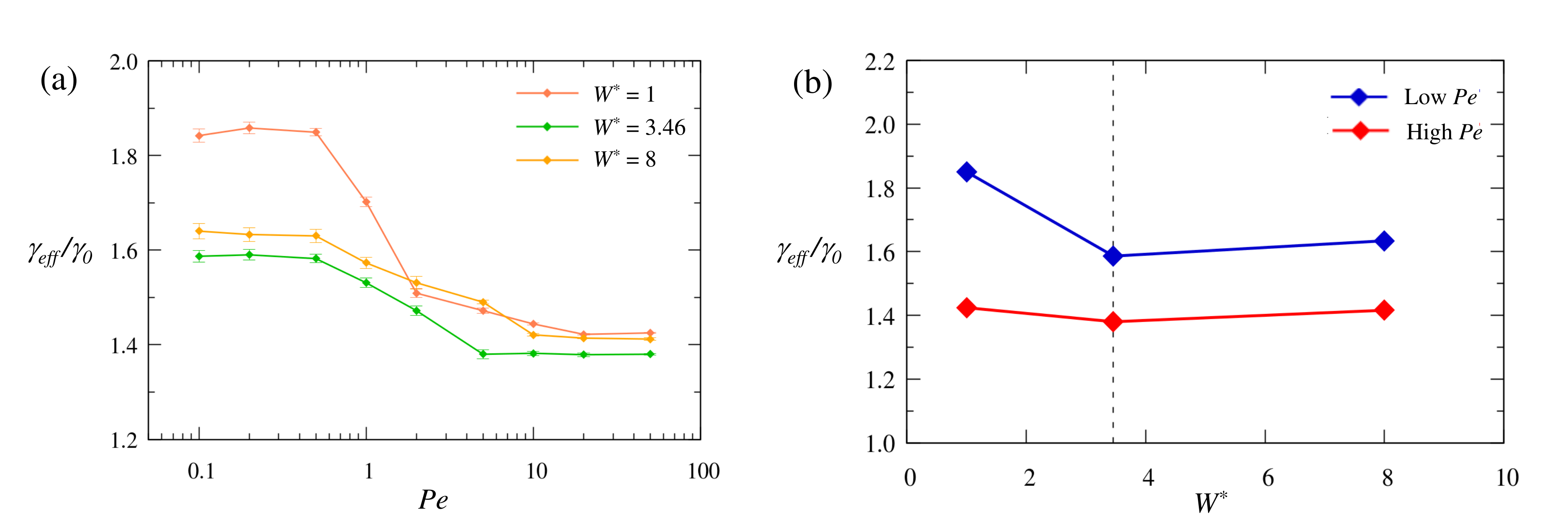}
   \caption{(a) Plot of $\gamma_{\rm eff}/\gamma_{0}$ \textit{vs.} $Pe$ for I phases at $\phi = 0.20$ at $W^{*} = 1$, $3.46$ and $W^{*} = 8$ (b) Variation of $\gamma_{\rm eff}/\gamma_{0}$ at high and low $Pe$ at different $W^{*}$ with $\phi = 0.20$. The dotted vertical line indicates the self-dual shape at $W^{*} = 3.46$. The lines are guides for the eye.}
    \label{Ch6_Fig2}
\end{figure*}

In Fig.\,\ref{Ch6_Fig1}(b), we show the dependence of the averaged effective friction coefficient, $\rm \gamma_{eff}/\gamma_{0}$, on the packing fraction. The upper curve has been obtained by averaging the friction coefficients calculated at $Pe \leq 0.5$, whereas the lower curve results from the average of friction coefficients measured in the high-$Pe$ regime. In agreement with past works \cite{puertas2014microrheology, daza2020dynamic}, $\rm \gamma_{eff}/\gamma_{0}$ tends to be larger in denser systems at both high and low $Pe$. This is somehow expected as the structure of denser systems is less prone to be distorted, hampering the tracer's mobility and resulting in larger effective frictions. Since HI are disregarded, the effect of $\phi$ on the effective friction coefficient of dual-shaped HBPs agrees well with the tendencies observed in past works that made similar assumptions \cite{carpen2005microrheology}.

\begin{figure*}[htbp!]
\centering
    \includegraphics[width=0.75\linewidth,height=0.61\textheight]{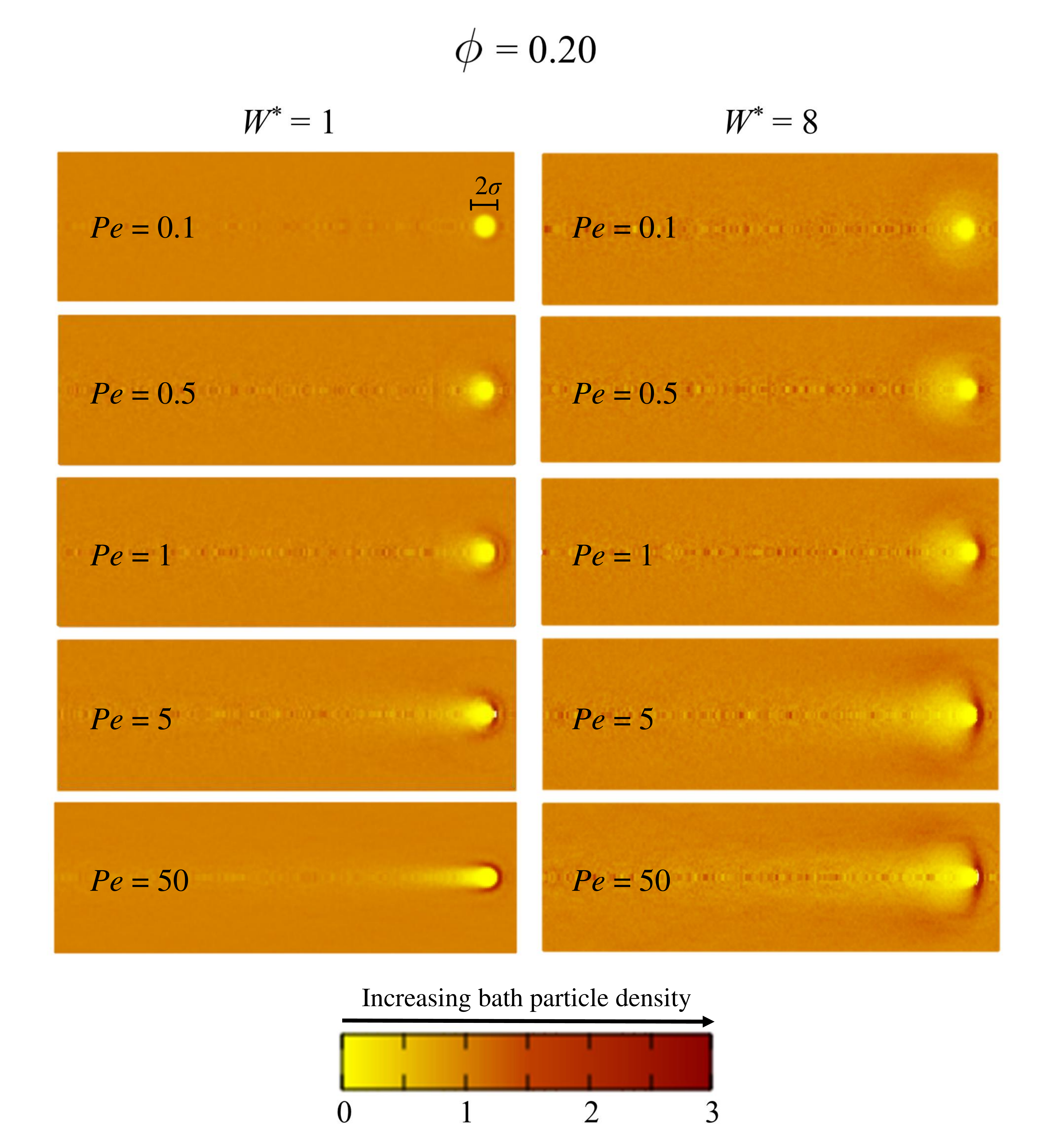}
    \caption{Density maps of HBPs with $\phi = 0.20$ at $W^{*} = 1$ (left) and $W^{*} = 8$ (right) and at the Pe numbers indicated in each frame. The colour palette is shown on at the bottom of the figure and refers to the ratio between the local density and the bath density. Yellow regions indicate low bath particle density, while the dark red regions indicate high bath particle density.}
    \label{Ch6_Fig4}
\end{figure*}

In light of these considerations, we now examine the effect of altering the shape anisotropy of HBPs on the effective friction induced by a tracer with diameter $\sigma = T$ at $\phi = 0.20$. In particular, we investigate the impact of particle anisotropy in baths of prolate ($W^{*} = 1$), self-dual-shaped ($W^{*} = 3.46$) and oblate ($W^{*} = 8$) HBPs. In Fig.\,\ref{Ch6_Fig2}(a), we report the $\gamma_{\rm eff}/\gamma_{0}$ \textit{vs} $Pe$ profile of different geometries. In all geometries studied, we once again observe three regimes associated with the effective friction: two plateau regimes corresponding to low and high $Pe$ numbers, and a force-thinning regime at intermediate $Pe$ numbers. In the low $Pe$ range ($0.1 \leq Pe \leq 0.5$), the effective friction coefficients for all $W^{*}$ are substantially constant, within statistical uncertainty. In this regime, the force applied to the tracer is too weak to perturb the microstructure of the HBP bath, so a symmetrical distribution of HBPs is found around the tracer. This tendency can be appreciated by looking at the density maps in Fig.\,\ref{Ch6_Fig4}, particularly at $Pe = 0.1$ for both $W^{*} = 1$ and $8$. At $Pe = 1$, all three systems enter a force-thinning regime, which is characterised by a reduction in the effective friction coefficient. The nonlinear regime of prolate HBPs ($W^{*} = 1$) spans in the range of $1 \leq Pe \leq 20$, while that of self-dual-shaped and oblate HBPs spans $1 \leq Pe \leq 5$ and $1 \leq Pe \leq 10$, respectively. Force-thinning occurs because the external force is strong enough to induce a microstructural distortion of the bath of HBPs, increasing its mobility. In the density maps of Fig.\,\ref{Ch6_Fig4}, we can see an asymmetry of the HBPs' density around the tracer with a region of high-density of HBPs in front of the tracer and low-density trail behind it for $Pe \geq 1$. At large $Pe$, the effective friction coefficients achieve a second plateau, which is due to a balance between the advection force from the tracer and a retarding force from the thin layer of very dense HBPs in front of the tracer that tends to scale proportionally with increasing $Pe$. As we can see clearly in Fig.\,\ref{Ch6_Fig2}(a), at large values of $Pe$, the effective friction coefficients of all geometries have very similar values.

Fig.\,\ref{Ch6_Fig2}(b) reports how the average effective friction coefficients vary with particle geometry. The low-$Pe$ effective friction coefficients are averaged across $0.1 \leq Pe \leq 0.5$ for all geometries. For high $Pe$, it is averaged from $Pe \geq 20$ for prolate HBPs, $Pe \geq 5$ for self-dual HBPs and $Pe \geq 10$ for oblate HBPs. From both Fig.\,\ref{Ch6_Fig2}(a) and Fig.\,\ref{Ch6_Fig2}(b), it is clear that the effective friction coefficient has a dependence on the geometry of the bath HBPs. At low $Pe$, we observe that $\rm {\gamma_{eff}/\gamma}_{0,W^{*} = 1} > \rm {\gamma_{eff}/\gamma}_{0,W^{*} = 8} > {\gamma_{eff}/\gamma}_{0,W^{*} = 3.46}$. At intermediate $Pe$ numbers, we observe crossovers between the effective friction coefficients at $W^{*} = 1$ and 8. Finally, in the high-$Pe$ regime, the effective friction coefficients have relatively similar values for the three particle geometries. We believe that these variations are due to an interplay between two factors: (\textit{i}) the relative size of the tracer with respect to that of HBPs and (\textit{ii}) the presence of nematic-like clusters at $\phi = 0.20$. The relative HBP/tracer size sets to what extent the tracer can perturb the microstructure of the bath. Due to the smaller surface area of prolate HBPs, one can assume that it is easier for the tracer to push away prolate HBPs and gain more mobility as compared to oblate HBPs. We note that our model particles do not possess an explicit mass. However, this can still be inferred from their diffusion coefficients at infinite dilution, whose magnitude goes with the inverse of a relevant characteristic length of the particle, which is related to its volume and thus to its mass. Oblate particles diffuse generally slower than prolate particles, except along the vector $\hat{\textbf{x}}$, as in this case the relevant surface area is $LT$, being the same for both oblate and prolate HBPs. The resistance to flow offered by larger surface areas can explain why the force-thinning of prolate HBPs is much more drastic in Fig\,\ref{Ch6_Fig2}(a). When the HBPs have an oblate anisotropy ($W^*=8$), it is harder for the tracer to push them away, so the tracer's mobility is only mildly enhanced with increasing $Pe$ and a relatively soft force-thinning is detected. In the density map of Fig.\,\ref{Ch6_Fig4}, one can observe that the low-density wake of $W^*=1$ at $Pe=50$ is approximately as large as the tracer diameter. This might be due to the rod-like shape of the bath particles, which can better accommodate around the tracer when perturbed, in contrast to suspensions of plate-like bath particles, whose low-density wake appears to be much larger than the tracer diameter. Since the cross-sectional area of the oblate particles is larger, there is a high likelihood for the tracer to be pushing the broader surface of the HBP, and this creates a larger low-density wake.

In addition to relative tracer/HBP size, also the phase behaviour of each system at $\phi = 0.20$ plays a role in determining the effective friction coefficient of the bath. Although all systems at this packing fraction are in the I phase, the features of their phase behaviour are different due to how far they are from their respective I-N phase boundary. In particular, at $\phi = 0.20$, the systems with prolate ($W^*=1$) and oblate ($W^*=8$) HBPs are closer to their respective I-N phase boundary than those made of self-dual-shaped HBPs ($W^*=3.46$) \cite{cuetos2017phase}. The packing fraction at which the I phase first transforms into a N phase is reported in Table\,\ref{Ch6_Tab1} for each value of the particle width \cite{cuetos2017phase}. We also report the difference, $\Delta \phi_{\rm I-N} \equiv \phi_{\rm I-N}-\phi$, between this transition point and the actual system packing, which is $\phi=0.20$ for the three geometries. We stress that the I-N transition in systems of hard cuboids has a strong first-order signature and what we are reporting in Table\,\ref{Ch6_Tab1} is the packing fraction at which the I phase must be compressed to transform into a N phase.

\begin{center}
\begin{table} [htbp]
	\centering
	\setlength{\tabcolsep}{14pt} 
	\renewcommand{\arraystretch}{1.2} 
	\begin{tabular}{c c c c c c c}
		\hline 
		\hline
		$W^{*}$ &  $\phi_{\rm I-N}$ & $\Delta \phi_{\rm I-N}$\\
		\hline
		\hline
		1.00 & 0.235 & $0.035$ \\
		3.46 &  0.319 & $0.119$ \\
		8.00 & 0.222 & $0.022$ \\
		\hline
		\hline%
		\end{tabular}
\caption{Difference in packing fraction of HBPs ($\Delta\phi_{\rm I-N}$) with $W^{*} = 1, 3.46$ and $8$ at $\phi = 0.20$ with their packing at their respective I-N phase boundary ($\phi_{\rm I-N}$). Here, $\Delta \phi_{\rm I-N} \equiv \phi_{\rm I-N}-\phi$ where the values of $\phi_{\rm I-N}$ are selected at the point where the I phase transitions into the N phase for each geometry.}
	\label{Ch6_Tab1}
\end{table}
\end{center}

We also notice that recent simulations in our group reported the formation of nematic-like clusters in systems of HBPs with $W^{*} = 1$ and 8 at $\phi = 0.20$, that is sufficiently close to the I-N phase transition \cite{tonti2021diffusion}. The presence of clusters of oriented cuboids, whose local packing can be larger than that of the surrounding I fluid, can slow down the tracer mobility and ultimately determine the MR of the whole system. In support of these arguments, the MR experiments by Paladugu and co-workers showed that N phases of bent-core mesogens exhibit unusual properties due to the presence of cybotactic (smectic) clusters \cite{paladugu2020microrheology}. These authors found that the viscosity anisotropy, defined as the difference between the viscosity measured along the nematic director and that perpendicular to it, is negative in cluster-free N phases and positive in the presence of cybotactic clusters. The increase of viscosity in the direction of the director, and the consequent reduction in particle self-diffusion, is most likely due to the compactness and packing of the oriented clusters.\\


\section{Conclusions}
 
In summary, we performed fixed-force active microrheology DMC simulations \cite{daza2020dynamic} to study the rheology of I fluids of colloidal cuboids. To gain an insight into their microrheological response, we computed the effective friction coefficient at increasing values of the $Pe$ number, a dimensionless quantity that sets the advection-to-thermal force ratio. In particular, at very small $Pe$, thermal forces dominate and the tracer's motion is essentially Brownian, whereas at very large $Pe$, external forces dominate and the tracer's motion is basically unaffected by the thermal fluctuations of the surronding bath. Tuning the system packing fraction, from $\phi=0.20$ to 0.30, and the particle width, from $W^{*}=1$ to $W^{*}=8$, dramatically influences the morphology, phase behaviour and dynamics of these fluids and, ultimately, their microrheological response. To clarify the effect of varying the packing fraction, we studied self-dual shaped HBPs, whose width, thickness and length are such that $W=\sqrt{LT}$. These systems have been selected as they exhibit stable I phases up to $\phi=0.30$ \cite{cuetos2017phase}. We found that the effective friction coefficients exhibit two linear regimes at low and high $Pe$ numbers and a force-thinning regime at intermediate $Pe$, a characteristic observed in colloidal systems when HI are neglected \cite{carpen2005microrheology, squires2005simple}. We found that increasing system density causes an enhanced friction experienced by the tracer since denser systems tend to have stronger local rigidity that is harder to disrupt. When varying the particle geometry from prolate ($W^{*} = 1$) to oblate ($W^{*} = 8$) at a fixed $\phi$, we found that the effective friction coefficients tend to show non-monotonic trends in all detected regimes. We believe that this behaviour could result from an interplay between the relative tracer/HBP size, and the presence of nematic-like clusters. At the moment, how these effects come into play is not fully clear, and more work is currently under consideration to fully understand these preliminary observations.

\section{Acknowledgements}

E.M.R. thanks the Malaysian Government Agency
Majlis Amanah Rakyat for funding his Ph.D. at the University of Manchester. A.P., L.T. and F.A.G.D. acknowledge financial support from the Leverhulme Trust Research Project Grant No.\,RPG-2018-415. A.P. is supported by a Maria Zambrano Senior distinguished researcher fellowship, financed by the European Union within the NextGenerationEU program. All authors acknowledge the assistance given by IT Services and the use of the Computational Shared Facility at the University of Manchester.

\bibliographystyle{apsrev4-1}
\bibliography{References.bib}

\begin{thebibliography}{61}%
\makeatletter
\providecommand \@ifxundefined [1]{%
 \@ifx{#1\undefined}
}%
\providecommand \@ifnum [1]{%
 \ifnum #1\expandafter \@firstoftwo
 \else \expandafter \@secondoftwo
 \fi
}%
\providecommand \@ifx [1]{%
 \ifx #1\expandafter \@firstoftwo
 \else \expandafter \@secondoftwo
 \fi
}%
\providecommand \natexlab [1]{#1}%
\providecommand \enquote  [1]{``#1''}%
\providecommand \bibnamefont  [1]{#1}%
\providecommand \bibfnamefont [1]{#1}%
\providecommand \citenamefont [1]{#1}%
\providecommand \href@noop [0]{\@secondoftwo}%
\providecommand \href [0]{\begingroup \@sanitize@url \@href}%
\providecommand \@href[1]{\@@startlink{#1}\@@href}%
\providecommand \@@href[1]{\endgroup#1\@@endlink}%
\providecommand \@sanitize@url [0]{\catcode `\\12\catcode `\$12\catcode
  `\&12\catcode `\#12\catcode `\^12\catcode `\_12\catcode `\%12\relax}%
\providecommand \@@startlink[1]{}%
\providecommand \@@endlink[0]{}%
\providecommand \url  [0]{\begingroup\@sanitize@url \@url }%
\providecommand \@url [1]{\endgroup\@href {#1}{\urlprefix }}%
\providecommand \urlprefix  [0]{URL }%
\providecommand \Eprint [0]{\href }%
\providecommand \doibase [0]{http://dx.doi.org/}%
\providecommand \selectlanguage [0]{\@gobble}%
\providecommand \bibinfo  [0]{\@secondoftwo}%
\providecommand \bibfield  [0]{\@secondoftwo}%
\providecommand \translation [1]{[#1]}%
\providecommand \BibitemOpen [0]{}%
\providecommand \bibitemStop [0]{}%
\providecommand \bibitemNoStop [0]{.\EOS\space}%
\providecommand \EOS [0]{\spacefactor3000\relax}%
\providecommand \BibitemShut  [1]{\csname bibitem#1\endcsname}%
\let\auto@bib@innerbib\@empty
\bibitem [{\citenamefont {Cicuta}\ and\ \citenamefont
  {Donald}(2007)}]{cicuta2007microrheology}%
  \BibitemOpen
  \bibfield  {author} {\bibinfo {author} {\bibfnamefont {P.}~\bibnamefont
  {Cicuta}}\ and\ \bibinfo {author} {\bibfnamefont {A.~M.}\ \bibnamefont
  {Donald}},\ }\href@noop {} {\bibfield  {journal} {\bibinfo  {journal} {Soft
  matter}\ }\textbf {\bibinfo {volume} {3}},\ \bibinfo {pages} {1449} (\bibinfo
  {year} {2007})}\BibitemShut {NoStop}%
\bibitem [{\citenamefont {Zia}(2018)}]{zia2018active}%
  \BibitemOpen
  \bibfield  {author} {\bibinfo {author} {\bibfnamefont {R.~N.}\ \bibnamefont
  {Zia}},\ }\href@noop {} {\bibfield  {journal} {\bibinfo  {journal} {Annual
  Review of Fluid Mechanics}\ }\textbf {\bibinfo {volume} {50}},\ \bibinfo
  {pages} {371} (\bibinfo {year} {2018})}\BibitemShut {NoStop}%
\bibitem [{\citenamefont {Habibi}\ \emph {et~al.}(2019)\citenamefont {Habibi},
  \citenamefont {Blanc}, \citenamefont {Mbarek},\ and\ \citenamefont
  {Soltani}}]{habibi2019passive}%
  \BibitemOpen
  \bibfield  {author} {\bibinfo {author} {\bibfnamefont {A.}~\bibnamefont
  {Habibi}}, \bibinfo {author} {\bibfnamefont {C.}~\bibnamefont {Blanc}},
  \bibinfo {author} {\bibfnamefont {N.~B.}\ \bibnamefont {Mbarek}}, \ and\
  \bibinfo {author} {\bibfnamefont {T.}~\bibnamefont {Soltani}},\ }\href@noop
  {} {\bibfield  {journal} {\bibinfo  {journal} {Journal of Molecular Liquids}\
  }\textbf {\bibinfo {volume} {288}},\ \bibinfo {pages} {111027} (\bibinfo
  {year} {2019})}\BibitemShut {NoStop}%
\bibitem [{\citenamefont {Chapman}\ and\ \citenamefont
  {Robertson-Anderson}(2014)}]{chapman2014nonlinear}%
  \BibitemOpen
  \bibfield  {author} {\bibinfo {author} {\bibfnamefont {C.~D.}\ \bibnamefont
  {Chapman}}\ and\ \bibinfo {author} {\bibfnamefont {R.~M.}\ \bibnamefont
  {Robertson-Anderson}},\ }\href@noop {} {\bibfield  {journal} {\bibinfo
  {journal} {Physical review letters}\ }\textbf {\bibinfo {volume} {113}},\
  \bibinfo {pages} {098303} (\bibinfo {year} {2014})}\BibitemShut {NoStop}%
\bibitem [{\citenamefont {Fernandez-Castanon}\ \emph
  {et~al.}(2018)\citenamefont {Fernandez-Castanon}, \citenamefont {Bianchi},
  \citenamefont {Saglimbeni}, \citenamefont {Di~Leonardo},\ and\ \citenamefont
  {Sciortino}}]{fernandez2018microrheology}%
  \BibitemOpen
  \bibfield  {author} {\bibinfo {author} {\bibfnamefont {J.}~\bibnamefont
  {Fernandez-Castanon}}, \bibinfo {author} {\bibfnamefont {S.}~\bibnamefont
  {Bianchi}}, \bibinfo {author} {\bibfnamefont {F.}~\bibnamefont {Saglimbeni}},
  \bibinfo {author} {\bibfnamefont {R.}~\bibnamefont {Di~Leonardo}}, \ and\
  \bibinfo {author} {\bibfnamefont {F.}~\bibnamefont {Sciortino}},\ }\href@noop
  {} {\bibfield  {journal} {\bibinfo  {journal} {Soft Matter}\ }\textbf
  {\bibinfo {volume} {14}},\ \bibinfo {pages} {6431} (\bibinfo {year}
  {2018})}\BibitemShut {NoStop}%
\bibitem [{\citenamefont {Levin}\ \emph {et~al.}(2020)\citenamefont {Levin},
  \citenamefont {Sorkin}, \citenamefont {Pine}, \citenamefont {Granek},
  \citenamefont {Bernheim-Groswasser},\ and\ \citenamefont
  {Roichman}}]{levin2020kinetics}%
  \BibitemOpen
  \bibfield  {author} {\bibinfo {author} {\bibfnamefont {M.}~\bibnamefont
  {Levin}}, \bibinfo {author} {\bibfnamefont {R.}~\bibnamefont {Sorkin}},
  \bibinfo {author} {\bibfnamefont {D.}~\bibnamefont {Pine}}, \bibinfo {author}
  {\bibfnamefont {R.}~\bibnamefont {Granek}}, \bibinfo {author} {\bibfnamefont
  {A.}~\bibnamefont {Bernheim-Groswasser}}, \ and\ \bibinfo {author}
  {\bibfnamefont {Y.}~\bibnamefont {Roichman}},\ }\href@noop {} {\bibfield
  {journal} {\bibinfo  {journal} {Soft matter}\ }\textbf {\bibinfo {volume}
  {16}},\ \bibinfo {pages} {7869} (\bibinfo {year} {2020})}\BibitemShut
  {NoStop}%
\bibitem [{\citenamefont {Weigand}\ \emph {et~al.}(2017)\citenamefont
  {Weigand}, \citenamefont {Messmore}, \citenamefont {Tu}, \citenamefont
  {Morales-Sanz}, \citenamefont {Blair}, \citenamefont {Deheyn}, \citenamefont
  {Urbach},\ and\ \citenamefont {Robertson-Anderson}}]{weigand2017active}%
  \BibitemOpen
  \bibfield  {author} {\bibinfo {author} {\bibfnamefont {W.}~\bibnamefont
  {Weigand}}, \bibinfo {author} {\bibfnamefont {A.}~\bibnamefont {Messmore}},
  \bibinfo {author} {\bibfnamefont {J.}~\bibnamefont {Tu}}, \bibinfo {author}
  {\bibfnamefont {A.}~\bibnamefont {Morales-Sanz}}, \bibinfo {author}
  {\bibfnamefont {D.}~\bibnamefont {Blair}}, \bibinfo {author} {\bibfnamefont
  {D.}~\bibnamefont {Deheyn}}, \bibinfo {author} {\bibfnamefont
  {J.}~\bibnamefont {Urbach}}, \ and\ \bibinfo {author} {\bibfnamefont
  {R.}~\bibnamefont {Robertson-Anderson}},\ }\href@noop {} {\bibfield
  {journal} {\bibinfo  {journal} {PloS one}\ }\textbf {\bibinfo {volume}
  {12}},\ \bibinfo {pages} {e0176732} (\bibinfo {year} {2017})}\BibitemShut
  {NoStop}%
\bibitem [{\citenamefont {Watts}\ \emph {et~al.}(2013)\citenamefont {Watts},
  \citenamefont {Tan}, \citenamefont {Wilson}, \citenamefont {Girkin},
  \citenamefont {Tassieri},\ and\ \citenamefont
  {Wright}}]{watts2013investigating}%
  \BibitemOpen
  \bibfield  {author} {\bibinfo {author} {\bibfnamefont {F.}~\bibnamefont
  {Watts}}, \bibinfo {author} {\bibfnamefont {L.~E.}\ \bibnamefont {Tan}},
  \bibinfo {author} {\bibfnamefont {C.~G.}\ \bibnamefont {Wilson}}, \bibinfo
  {author} {\bibfnamefont {J.~M.}\ \bibnamefont {Girkin}}, \bibinfo {author}
  {\bibfnamefont {M.}~\bibnamefont {Tassieri}}, \ and\ \bibinfo {author}
  {\bibfnamefont {A.~J.}\ \bibnamefont {Wright}},\ }\href@noop {} {\bibfield
  {journal} {\bibinfo  {journal} {Journal of Optics}\ }\textbf {\bibinfo
  {volume} {16}},\ \bibinfo {pages} {015301} (\bibinfo {year}
  {2013})}\BibitemShut {NoStop}%
\bibitem [{\citenamefont {Wilson}\ \emph {et~al.}(2009)\citenamefont {Wilson},
  \citenamefont {Harrison}, \citenamefont {Schofield}, \citenamefont {Arlt},\
  and\ \citenamefont {Poon}}]{wilson2009passive}%
  \BibitemOpen
  \bibfield  {author} {\bibinfo {author} {\bibfnamefont {L.}~\bibnamefont
  {Wilson}}, \bibinfo {author} {\bibfnamefont {A.}~\bibnamefont {Harrison}},
  \bibinfo {author} {\bibfnamefont {A.~B.}\ \bibnamefont {Schofield}}, \bibinfo
  {author} {\bibfnamefont {J.}~\bibnamefont {Arlt}}, \ and\ \bibinfo {author}
  {\bibfnamefont {W.}~\bibnamefont {Poon}},\ }\href@noop {} {\bibfield
  {journal} {\bibinfo  {journal} {The Journal of Physical Chemistry B}\
  }\textbf {\bibinfo {volume} {113}},\ \bibinfo {pages} {3806} (\bibinfo {year}
  {2009})}\BibitemShut {NoStop}%
\bibitem [{\citenamefont {Nazockdast}\ and\ \citenamefont
  {Morris}(2016)}]{Nazockdast2016}%
  \BibitemOpen
  \bibfield  {author} {\bibinfo {author} {\bibfnamefont {E.}~\bibnamefont
  {Nazockdast}}\ and\ \bibinfo {author} {\bibfnamefont {J.~F.}\ \bibnamefont
  {Morris}},\ }\href {\doibase 10.1122/1.4954201} {\bibfield  {journal}
  {\bibinfo  {journal} {Journal of Rheology}\ }\textbf {\bibinfo {volume}
  {60}},\ \bibinfo {pages} {733} (\bibinfo {year} {2016})}\BibitemShut
  {NoStop}%
\bibitem [{\citenamefont {Carpen}\ and\ \citenamefont
  {Brady}(2005)}]{carpen2005microrheology}%
  \BibitemOpen
  \bibfield  {author} {\bibinfo {author} {\bibfnamefont {I.~C.}\ \bibnamefont
  {Carpen}}\ and\ \bibinfo {author} {\bibfnamefont {J.~F.}\ \bibnamefont
  {Brady}},\ }\href@noop {} {\bibfield  {journal} {\bibinfo  {journal} {Journal
  of Rheology}\ }\textbf {\bibinfo {volume} {49}},\ \bibinfo {pages} {1483}
  (\bibinfo {year} {2005})}\BibitemShut {NoStop}%
\bibitem [{\citenamefont {Dussi}\ \emph {et~al.}(2018)\citenamefont {Dussi},
  \citenamefont {Tasios}, \citenamefont {Drwenski}, \citenamefont {van Roij},\
  and\ \citenamefont {Dijkstra}}]{dussi2018hard}%
  \BibitemOpen
  \bibfield  {author} {\bibinfo {author} {\bibfnamefont {S.}~\bibnamefont
  {Dussi}}, \bibinfo {author} {\bibfnamefont {N.}~\bibnamefont {Tasios}},
  \bibinfo {author} {\bibfnamefont {T.}~\bibnamefont {Drwenski}}, \bibinfo
  {author} {\bibfnamefont {R.}~\bibnamefont {van Roij}}, \ and\ \bibinfo
  {author} {\bibfnamefont {M.}~\bibnamefont {Dijkstra}},\ }\href@noop {}
  {\bibfield  {journal} {\bibinfo  {journal} {Phys. Rev. Lett.}\ }\textbf
  {\bibinfo {volume} {120}},\ \bibinfo {pages} {177801} (\bibinfo {year}
  {2018})}\BibitemShut {NoStop}%
\bibitem [{\citenamefont {Tasios}\ and\ \citenamefont
  {Dijkstra}(2017)}]{tasios2017simulation}%
  \BibitemOpen
  \bibfield  {author} {\bibinfo {author} {\bibfnamefont {N.}~\bibnamefont
  {Tasios}}\ and\ \bibinfo {author} {\bibfnamefont {M.}~\bibnamefont
  {Dijkstra}},\ }\href@noop {} {\bibfield  {journal} {\bibinfo  {journal} {J.
  Chem. Phys.}\ }\textbf {\bibinfo {volume} {146}},\ \bibinfo {pages} {144901}
  (\bibinfo {year} {2017})}\BibitemShut {NoStop}%
\bibitem [{\citenamefont {Chiappini}\ \emph {et~al.}(2019)\citenamefont
  {Chiappini}, \citenamefont {Drwenski}, \citenamefont {van Roij},\ and\
  \citenamefont {Dijkstra}}]{chiappini2019biaxial}%
  \BibitemOpen
  \bibfield  {author} {\bibinfo {author} {\bibfnamefont {M.}~\bibnamefont
  {Chiappini}}, \bibinfo {author} {\bibfnamefont {T.}~\bibnamefont {Drwenski}},
  \bibinfo {author} {\bibfnamefont {R.}~\bibnamefont {van Roij}}, \ and\
  \bibinfo {author} {\bibfnamefont {M.}~\bibnamefont {Dijkstra}},\ }\href@noop
  {} {\bibfield  {journal} {\bibinfo  {journal} {Phys. Rev. Lett.}\ }\textbf
  {\bibinfo {volume} {123}},\ \bibinfo {pages} {068001} (\bibinfo {year}
  {2019})}\BibitemShut {NoStop}%
\bibitem [{\citenamefont {Fern{\'a}ndez-Rico}\ \emph
  {et~al.}(2020)\citenamefont {Fern{\'a}ndez-Rico}, \citenamefont {Chiappini},
  \citenamefont {Yanagishima}, \citenamefont {de~Sousa}, \citenamefont {Aarts},
  \citenamefont {Dijkstra},\ and\ \citenamefont
  {Dullens}}]{fernandez2020shaping}%
  \BibitemOpen
  \bibfield  {author} {\bibinfo {author} {\bibfnamefont {C.}~\bibnamefont
  {Fern{\'a}ndez-Rico}}, \bibinfo {author} {\bibfnamefont {M.}~\bibnamefont
  {Chiappini}}, \bibinfo {author} {\bibfnamefont {T.}~\bibnamefont
  {Yanagishima}}, \bibinfo {author} {\bibfnamefont {H.}~\bibnamefont
  {de~Sousa}}, \bibinfo {author} {\bibfnamefont {D.~G.}\ \bibnamefont {Aarts}},
  \bibinfo {author} {\bibfnamefont {M.}~\bibnamefont {Dijkstra}}, \ and\
  \bibinfo {author} {\bibfnamefont {R.~P.}\ \bibnamefont {Dullens}},\
  }\href@noop {} {\bibfield  {journal} {\bibinfo  {journal} {Science}\ }\textbf
  {\bibinfo {volume} {369}},\ \bibinfo {pages} {950} (\bibinfo {year}
  {2020})}\BibitemShut {NoStop}%
\bibitem [{\citenamefont {Berardi}\ and\ \citenamefont
  {Zannoni}(2000)}]{berardi2000thermotropic}%
  \BibitemOpen
  \bibfield  {author} {\bibinfo {author} {\bibfnamefont {R.}~\bibnamefont
  {Berardi}}\ and\ \bibinfo {author} {\bibfnamefont {C.}~\bibnamefont
  {Zannoni}},\ }\href@noop {} {\bibfield  {journal} {\bibinfo  {journal} {The
  Journal of Chemical Physics}\ }\textbf {\bibinfo {volume} {113}},\ \bibinfo
  {pages} {5971} (\bibinfo {year} {2000})}\BibitemShut {NoStop}%
\bibitem [{\citenamefont {Lehmann}\ \emph {et~al.}(2019)\citenamefont
  {Lehmann}, \citenamefont {Maisch}, \citenamefont {Scheuring}, \citenamefont
  {Carvalho}, \citenamefont {Cruz}, \citenamefont {Sebasti{\~a}o},\ and\
  \citenamefont {Dong}}]{lehmann2019molecular}%
  \BibitemOpen
  \bibfield  {author} {\bibinfo {author} {\bibfnamefont {M.}~\bibnamefont
  {Lehmann}}, \bibinfo {author} {\bibfnamefont {S.}~\bibnamefont {Maisch}},
  \bibinfo {author} {\bibfnamefont {N.}~\bibnamefont {Scheuring}}, \bibinfo
  {author} {\bibfnamefont {J.}~\bibnamefont {Carvalho}}, \bibinfo {author}
  {\bibfnamefont {C.}~\bibnamefont {Cruz}}, \bibinfo {author} {\bibfnamefont
  {P.~J.}\ \bibnamefont {Sebasti{\~a}o}}, \ and\ \bibinfo {author}
  {\bibfnamefont {R.~Y.}\ \bibnamefont {Dong}},\ }\href@noop {} {\bibfield
  {journal} {\bibinfo  {journal} {Soft matter}\ }\textbf {\bibinfo {volume}
  {15}},\ \bibinfo {pages} {8496} (\bibinfo {year} {2019})}\BibitemShut
  {NoStop}%
\bibitem [{\citenamefont {van~den Pol}\ \emph {et~al.}(2009)\citenamefont
  {van~den Pol}, \citenamefont {Petukhov}, \citenamefont {Thies-Weesie},
  \citenamefont {Byelov},\ and\ \citenamefont {Vroege}}]{van2009experimental}%
  \BibitemOpen
  \bibfield  {author} {\bibinfo {author} {\bibfnamefont {E.}~\bibnamefont
  {van~den Pol}}, \bibinfo {author} {\bibfnamefont {A.~V.}\ \bibnamefont
  {Petukhov}}, \bibinfo {author} {\bibfnamefont {D.~M.~E.}\ \bibnamefont
  {Thies-Weesie}}, \bibinfo {author} {\bibfnamefont {D.~V.}\ \bibnamefont
  {Byelov}}, \ and\ \bibinfo {author} {\bibfnamefont {G.~J.}\ \bibnamefont
  {Vroege}},\ }\href@noop {} {\bibfield  {journal} {\bibinfo  {journal} {Phys.
  Rev. Lett.}\ }\textbf {\bibinfo {volume} {103}},\ \bibinfo {pages} {258301}
  (\bibinfo {year} {2009})}\BibitemShut {NoStop}%
\bibitem [{\citenamefont {van~den Pol}\ \emph
  {et~al.}(2010{\natexlab{a}})\citenamefont {van~den Pol}, \citenamefont
  {Lupascu}, \citenamefont {Diaconeasa}, \citenamefont {Petukhov},
  \citenamefont {Byelov},\ and\ \citenamefont {Vroege}}]{van2010onsager}%
  \BibitemOpen
  \bibfield  {author} {\bibinfo {author} {\bibfnamefont {E.}~\bibnamefont
  {van~den Pol}}, \bibinfo {author} {\bibfnamefont {A.}~\bibnamefont
  {Lupascu}}, \bibinfo {author} {\bibfnamefont {M.}~\bibnamefont {Diaconeasa}},
  \bibinfo {author} {\bibfnamefont {A.}~\bibnamefont {Petukhov}}, \bibinfo
  {author} {\bibfnamefont {D.}~\bibnamefont {Byelov}}, \ and\ \bibinfo {author}
  {\bibfnamefont {G.}~\bibnamefont {Vroege}},\ }\href@noop {} {\bibfield
  {journal} {\bibinfo  {journal} {Chem. Phys. Lett.}\ }\textbf {\bibinfo
  {volume} {1}},\ \bibinfo {pages} {2174} (\bibinfo {year}
  {2010}{\natexlab{a}})}\BibitemShut {NoStop}%
\bibitem [{\citenamefont {van~den Pol}\ \emph
  {et~al.}(2010{\natexlab{b}})\citenamefont {van~den Pol}, \citenamefont
  {Lupascu}, \citenamefont {Davidson},\ and\ \citenamefont
  {Vroege}}]{van2010isotropic}%
  \BibitemOpen
  \bibfield  {author} {\bibinfo {author} {\bibfnamefont {E.}~\bibnamefont
  {van~den Pol}}, \bibinfo {author} {\bibfnamefont {A.}~\bibnamefont
  {Lupascu}}, \bibinfo {author} {\bibfnamefont {P.}~\bibnamefont {Davidson}}, \
  and\ \bibinfo {author} {\bibfnamefont {G.}~\bibnamefont {Vroege}},\
  }\href@noop {} {\bibfield  {journal} {\bibinfo  {journal} {J. Chem. Phys.}\
  }\textbf {\bibinfo {volume} {133}},\ \bibinfo {pages} {164504} (\bibinfo
  {year} {2010}{\natexlab{b}})}\BibitemShut {NoStop}%
\bibitem [{\citenamefont {Mart{\'\i}nez-Rat{\'o}n}\ \emph
  {et~al.}(2011)\citenamefont {Mart{\'\i}nez-Rat{\'o}n}, \citenamefont
  {Varga},\ and\ \citenamefont {Velasco}}]{martinez2011biaxial}%
  \BibitemOpen
  \bibfield  {author} {\bibinfo {author} {\bibfnamefont {Y.}~\bibnamefont
  {Mart{\'\i}nez-Rat{\'o}n}}, \bibinfo {author} {\bibfnamefont
  {S.}~\bibnamefont {Varga}}, \ and\ \bibinfo {author} {\bibfnamefont
  {E.}~\bibnamefont {Velasco}},\ }\href@noop {} {\bibfield  {journal} {\bibinfo
   {journal} {Phys. Chem. Chem. Phys.}\ }\textbf {\bibinfo {volume} {13}},\
  \bibinfo {pages} {13247} (\bibinfo {year} {2011})}\BibitemShut {NoStop}%
\bibitem [{\citenamefont {Belli}\ \emph {et~al.}(2011)\citenamefont {Belli},
  \citenamefont {Patti}, \citenamefont {Dijkstra},\ and\ \citenamefont {van
  Roij}}]{belli2011polydispersity}%
  \BibitemOpen
  \bibfield  {author} {\bibinfo {author} {\bibfnamefont {S.}~\bibnamefont
  {Belli}}, \bibinfo {author} {\bibfnamefont {A.}~\bibnamefont {Patti}},
  \bibinfo {author} {\bibfnamefont {M.}~\bibnamefont {Dijkstra}}, \ and\
  \bibinfo {author} {\bibfnamefont {R.}~\bibnamefont {van Roij}},\ }\href@noop
  {} {\bibfield  {journal} {\bibinfo  {journal} {Phys. Rev. Lett.}\ }\textbf
  {\bibinfo {volume} {107}},\ \bibinfo {pages} {148303} (\bibinfo {year}
  {2011})}\BibitemShut {NoStop}%
\bibitem [{\citenamefont {Belli}\ \emph {et~al.}(2012)\citenamefont {Belli},
  \citenamefont {Dijkstra},\ and\ \citenamefont {van
  Roij}}]{belli2012depletion}%
  \BibitemOpen
  \bibfield  {author} {\bibinfo {author} {\bibfnamefont {S.}~\bibnamefont
  {Belli}}, \bibinfo {author} {\bibfnamefont {M.}~\bibnamefont {Dijkstra}}, \
  and\ \bibinfo {author} {\bibfnamefont {R.}~\bibnamefont {van Roij}},\
  }\href@noop {} {\bibfield  {journal} {\bibinfo  {journal} {J. Phys.: Condens.
  Matter}\ }\textbf {\bibinfo {volume} {24}},\ \bibinfo {pages} {284128}
  (\bibinfo {year} {2012})}\BibitemShut {NoStop}%
\bibitem [{\citenamefont {Peroukidis}\ and\ \citenamefont
  {Vanakaras}(2013)}]{peroukidis2013phase}%
  \BibitemOpen
  \bibfield  {author} {\bibinfo {author} {\bibfnamefont {S.~D.}\ \bibnamefont
  {Peroukidis}}\ and\ \bibinfo {author} {\bibfnamefont {A.~G.}\ \bibnamefont
  {Vanakaras}},\ }\href@noop {} {\bibfield  {journal} {\bibinfo  {journal}
  {Soft Matter}\ }\textbf {\bibinfo {volume} {9}},\ \bibinfo {pages} {7419}
  (\bibinfo {year} {2013})}\BibitemShut {NoStop}%
\bibitem [{\citenamefont {Peroukidis}\ \emph {et~al.}(2013)\citenamefont
  {Peroukidis}, \citenamefont {Vanakaras},\ and\ \citenamefont
  {Photinos}}]{peroukidis2013supramolecular}%
  \BibitemOpen
  \bibfield  {author} {\bibinfo {author} {\bibfnamefont {S.~D.}\ \bibnamefont
  {Peroukidis}}, \bibinfo {author} {\bibfnamefont {A.~G.}\ \bibnamefont
  {Vanakaras}}, \ and\ \bibinfo {author} {\bibfnamefont {D.~J.}\ \bibnamefont
  {Photinos}},\ }\href@noop {} {\bibfield  {journal} {\bibinfo  {journal}
  {Phys. Rev. E}\ }\textbf {\bibinfo {volume} {88}},\ \bibinfo {pages} {062508}
  (\bibinfo {year} {2013})}\BibitemShut {NoStop}%
\bibitem [{\citenamefont {Peroukidis}(2014)}]{peroukidis2014biaxial}%
  \BibitemOpen
  \bibfield  {author} {\bibinfo {author} {\bibfnamefont {S.~D.}\ \bibnamefont
  {Peroukidis}},\ }\href@noop {} {\bibfield  {journal} {\bibinfo  {journal}
  {Soft Matter}\ }\textbf {\bibinfo {volume} {10}},\ \bibinfo {pages} {4199}
  (\bibinfo {year} {2014})}\BibitemShut {NoStop}%
\bibitem [{\citenamefont {Mederos}\ \emph {et~al.}(2014)\citenamefont
  {Mederos}, \citenamefont {Velasco},\ and\ \citenamefont
  {Mart{\'\i}nez-Rat{\'o}n}}]{mederos2014hard}%
  \BibitemOpen
  \bibfield  {author} {\bibinfo {author} {\bibfnamefont {L.}~\bibnamefont
  {Mederos}}, \bibinfo {author} {\bibfnamefont {E.}~\bibnamefont {Velasco}}, \
  and\ \bibinfo {author} {\bibfnamefont {Y.}~\bibnamefont
  {Mart{\'\i}nez-Rat{\'o}n}},\ }\href@noop {} {\bibfield  {journal} {\bibinfo
  {journal} {J. Phys.: Condens. Matter}\ }\textbf {\bibinfo {volume} {26}},\
  \bibinfo {pages} {463101} (\bibinfo {year} {2014})}\BibitemShut {NoStop}%
\bibitem [{\citenamefont {Gonz{\'a}lez-Pinto}\ \emph
  {et~al.}(2015)\citenamefont {Gonz{\'a}lez-Pinto}, \citenamefont
  {Mart{\'\i}nez-Rat{\'o}n}, \citenamefont {Velasco},\ and\ \citenamefont
  {Varga}}]{gonzalez2015effect}%
  \BibitemOpen
  \bibfield  {author} {\bibinfo {author} {\bibfnamefont {M.}~\bibnamefont
  {Gonz{\'a}lez-Pinto}}, \bibinfo {author} {\bibfnamefont {Y.}~\bibnamefont
  {Mart{\'\i}nez-Rat{\'o}n}}, \bibinfo {author} {\bibfnamefont
  {E.}~\bibnamefont {Velasco}}, \ and\ \bibinfo {author} {\bibfnamefont
  {S.}~\bibnamefont {Varga}},\ }\href@noop {} {\bibfield  {journal} {\bibinfo
  {journal} {Phys. Chem. Chem. Phys.}\ }\textbf {\bibinfo {volume} {17}},\
  \bibinfo {pages} {6389} (\bibinfo {year} {2015})}\BibitemShut {NoStop}%
\bibitem [{\citenamefont {Cuetos}\ \emph {et~al.}(2017)\citenamefont {Cuetos},
  \citenamefont {Dennison}, \citenamefont {Masters},\ and\ \citenamefont
  {Patti}}]{cuetos2017phase}%
  \BibitemOpen
  \bibfield  {author} {\bibinfo {author} {\bibfnamefont {A.}~\bibnamefont
  {Cuetos}}, \bibinfo {author} {\bibfnamefont {M.}~\bibnamefont {Dennison}},
  \bibinfo {author} {\bibfnamefont {A.}~\bibnamefont {Masters}}, \ and\
  \bibinfo {author} {\bibfnamefont {A.}~\bibnamefont {Patti}},\ }\href@noop {}
  {\bibfield  {journal} {\bibinfo  {journal} {Soft Matter}\ }\textbf {\bibinfo
  {volume} {13}},\ \bibinfo {pages} {4720} (\bibinfo {year}
  {2017})}\BibitemShut {NoStop}%
\bibitem [{\citenamefont {Yang}\ \emph {et~al.}(2018)\citenamefont {Yang},
  \citenamefont {Chen}, \citenamefont {Thanneeru}, \citenamefont {He},
  \citenamefont {Liu},\ and\ \citenamefont {Nie}}]{yang2018synthesis}%
  \BibitemOpen
  \bibfield  {author} {\bibinfo {author} {\bibfnamefont {Y.}~\bibnamefont
  {Yang}}, \bibinfo {author} {\bibfnamefont {G.}~\bibnamefont {Chen}}, \bibinfo
  {author} {\bibfnamefont {S.}~\bibnamefont {Thanneeru}}, \bibinfo {author}
  {\bibfnamefont {J.}~\bibnamefont {He}}, \bibinfo {author} {\bibfnamefont
  {K.}~\bibnamefont {Liu}}, \ and\ \bibinfo {author} {\bibfnamefont
  {Z.}~\bibnamefont {Nie}},\ }\href@noop {} {\bibfield  {journal} {\bibinfo
  {journal} {Nat. Commun.}\ }\textbf {\bibinfo {volume} {9}},\ \bibinfo {pages}
  {4513} (\bibinfo {year} {2018})}\BibitemShut {NoStop}%
\bibitem [{\citenamefont {Patti}\ and\ \citenamefont
  {Cuetos}(2018)}]{patti2018monte}%
  \BibitemOpen
  \bibfield  {author} {\bibinfo {author} {\bibfnamefont {A.}~\bibnamefont
  {Patti}}\ and\ \bibinfo {author} {\bibfnamefont {A.}~\bibnamefont {Cuetos}},\
  }\href@noop {} {\bibfield  {journal} {\bibinfo  {journal} {Mol. Simul.}\
  }\textbf {\bibinfo {volume} {44}},\ \bibinfo {pages} {516} (\bibinfo {year}
  {2018})}\BibitemShut {NoStop}%
\bibitem [{\citenamefont {Cuetos}\ \emph {et~al.}(2019)\citenamefont {Cuetos},
  \citenamefont {Mirzad~Rafael}, \citenamefont {Corbett},\ and\ \citenamefont
  {Patti}}]{cuetos2019biaxial}%
  \BibitemOpen
  \bibfield  {author} {\bibinfo {author} {\bibfnamefont {A.}~\bibnamefont
  {Cuetos}}, \bibinfo {author} {\bibfnamefont {E.}~\bibnamefont
  {Mirzad~Rafael}}, \bibinfo {author} {\bibfnamefont {D.}~\bibnamefont
  {Corbett}}, \ and\ \bibinfo {author} {\bibfnamefont {A.}~\bibnamefont
  {Patti}},\ }\href@noop {} {\bibfield  {journal} {\bibinfo  {journal} {Soft
  Matter}\ }\textbf {\bibinfo {volume} {15}},\ \bibinfo {pages} {1922}
  (\bibinfo {year} {2019})}\BibitemShut {NoStop}%
\bibitem [{\citenamefont {Mirzad~Rafael}\ \emph {et~al.}(2020)\citenamefont
  {Mirzad~Rafael}, \citenamefont {Cuetos}, \citenamefont {Corbett},\ and\
  \citenamefont {Patti}}]{rafael2020self}%
  \BibitemOpen
  \bibfield  {author} {\bibinfo {author} {\bibfnamefont {E.}~\bibnamefont
  {Mirzad~Rafael}}, \bibinfo {author} {\bibfnamefont {A.}~\bibnamefont
  {Cuetos}}, \bibinfo {author} {\bibfnamefont {D.}~\bibnamefont {Corbett}}, \
  and\ \bibinfo {author} {\bibfnamefont {A.}~\bibnamefont {Patti}},\
  }\href@noop {} {\bibfield  {journal} {\bibinfo  {journal} {Soft Matter}\
  }\textbf {\bibinfo {volume} {16}},\ \bibinfo {pages} {5565} (\bibinfo {year}
  {2020})}\BibitemShut {NoStop}%
\bibitem [{\citenamefont {Skutnik}\ \emph {et~al.}(2020)\citenamefont
  {Skutnik}, \citenamefont {Geier},\ and\ \citenamefont
  {Schoen}}]{skutnik2020biaxial}%
  \BibitemOpen
  \bibfield  {author} {\bibinfo {author} {\bibfnamefont {R.~A.}\ \bibnamefont
  {Skutnik}}, \bibinfo {author} {\bibfnamefont {I.~S.}\ \bibnamefont {Geier}},
  \ and\ \bibinfo {author} {\bibfnamefont {M.}~\bibnamefont {Schoen}},\
  }\href@noop {} {\bibfield  {journal} {\bibinfo  {journal} {Mol. Phys.}\ ,\
  \bibinfo {pages} {1}} (\bibinfo {year} {2020})}\BibitemShut {NoStop}%
\bibitem [{\citenamefont {Cuetos}\ and\ \citenamefont
  {Patti}(2020)}]{cuetos2020dynamics}%
  \BibitemOpen
  \bibfield  {author} {\bibinfo {author} {\bibfnamefont {A.}~\bibnamefont
  {Cuetos}}\ and\ \bibinfo {author} {\bibfnamefont {A.}~\bibnamefont {Patti}},\
  }\href@noop {} {\bibfield  {journal} {\bibinfo  {journal} {Phys. Rev. E}\
  }\textbf {\bibinfo {volume} {101}},\ \bibinfo {pages} {052702} (\bibinfo
  {year} {2020})}\BibitemShut {NoStop}%
\bibitem [{\citenamefont {Mirzad~Rafael}\ \emph {et~al.}(2021)\citenamefont
  {Mirzad~Rafael}, \citenamefont {Tonti}, \citenamefont {Corbett},
  \citenamefont {Cuetos},\ and\ \citenamefont {Patti}}]{mirzad2021dynamics}%
  \BibitemOpen
  \bibfield  {author} {\bibinfo {author} {\bibfnamefont {E.}~\bibnamefont
  {Mirzad~Rafael}}, \bibinfo {author} {\bibfnamefont {L.}~\bibnamefont
  {Tonti}}, \bibinfo {author} {\bibfnamefont {D.}~\bibnamefont {Corbett}},
  \bibinfo {author} {\bibfnamefont {A.}~\bibnamefont {Cuetos}}, \ and\ \bibinfo
  {author} {\bibfnamefont {A.}~\bibnamefont {Patti}},\ }\href@noop {}
  {\bibfield  {journal} {\bibinfo  {journal} {Physics of Fluids}\ }\textbf
  {\bibinfo {volume} {33}},\ \bibinfo {pages} {067115} (\bibinfo {year}
  {2021})}\BibitemShut {NoStop}%
\bibitem [{\citenamefont {Patti}\ and\ \citenamefont
  {Cuetos}(2021)}]{patti2021dynamics}%
  \BibitemOpen
  \bibfield  {author} {\bibinfo {author} {\bibfnamefont {A.}~\bibnamefont
  {Patti}}\ and\ \bibinfo {author} {\bibfnamefont {A.}~\bibnamefont {Cuetos}},\
  }\href@noop {} {\bibfield  {journal} {\bibinfo  {journal} {Physics of
  Fluids}\ }\textbf {\bibinfo {volume} {33}},\ \bibinfo {pages} {097103}
  (\bibinfo {year} {2021})}\BibitemShut {NoStop}%
\bibitem [{\citenamefont {Gabriel}\ and\ \citenamefont
  {Davidson}(2000)}]{gabriel2000}%
  \BibitemOpen
  \bibfield  {author} {\bibinfo {author} {\bibfnamefont {J.-C.~P.}\
  \bibnamefont {Gabriel}}\ and\ \bibinfo {author} {\bibfnamefont
  {P.}~\bibnamefont {Davidson}},\ }\href {\doibase
  https://doi.org/10.1002/(SICI)1521-4095(200001)12:1<9::AID-ADMA9>3.0.CO;2-6}
  {\bibfield  {journal} {\bibinfo  {journal} {Advanced Materials}\ }\textbf
  {\bibinfo {volume} {12}},\ \bibinfo {pages} {9} (\bibinfo {year}
  {2000})}\BibitemShut {NoStop}%
\bibitem [{\citenamefont {Tschierske}\ and\ \citenamefont
  {Photinos}(2010)}]{tschierske2010biaxial}%
  \BibitemOpen
  \bibfield  {author} {\bibinfo {author} {\bibfnamefont {C.}~\bibnamefont
  {Tschierske}}\ and\ \bibinfo {author} {\bibfnamefont {D.~J.}\ \bibnamefont
  {Photinos}},\ }\href@noop {} {\bibfield  {journal} {\bibinfo  {journal}
  {Journal of Materials Chemistry}\ }\textbf {\bibinfo {volume} {20}},\
  \bibinfo {pages} {4263} (\bibinfo {year} {2010})}\BibitemShut {NoStop}%
\bibitem [{\citenamefont {Lekkerkerker}\ and\ \citenamefont
  {Vroege}(2013)}]{Lekkerkerker2013}%
  \BibitemOpen
  \bibfield  {author} {\bibinfo {author} {\bibfnamefont {H.~N.~W.}\
  \bibnamefont {Lekkerkerker}}\ and\ \bibinfo {author} {\bibfnamefont {G.~J.}\
  \bibnamefont {Vroege}},\ }\href {\doibase 10.1098/rsta.2012.0263} {\bibfield
  {journal} {\bibinfo  {journal} {Philosophical Transactions of the Royal
  Society A: Mathematical, Physical and Engineering Sciences}\ }\textbf
  {\bibinfo {volume} {371}},\ \bibinfo {pages} {20120263} (\bibinfo {year}
  {2013})}\BibitemShut {NoStop}%
\bibitem [{\citenamefont {Leferink~op Reinink}\ \emph
  {et~al.}(2014)\citenamefont {Leferink~op Reinink}, \citenamefont {Belli},
  \citenamefont {van Roij}, \citenamefont {Dijkstra}, \citenamefont
  {Petukhov},\ and\ \citenamefont {Vroege}}]{Leferink2014}%
  \BibitemOpen
  \bibfield  {author} {\bibinfo {author} {\bibfnamefont {A.~B. G.~M.}\
  \bibnamefont {Leferink~op Reinink}}, \bibinfo {author} {\bibfnamefont
  {S.}~\bibnamefont {Belli}}, \bibinfo {author} {\bibfnamefont
  {R.}~\bibnamefont {van Roij}}, \bibinfo {author} {\bibfnamefont
  {M.}~\bibnamefont {Dijkstra}}, \bibinfo {author} {\bibfnamefont {A.~V.}\
  \bibnamefont {Petukhov}}, \ and\ \bibinfo {author} {\bibfnamefont {G.~J.}\
  \bibnamefont {Vroege}},\ }\href {\doibase 10.1039/C3SM52242C} {\bibfield
  {journal} {\bibinfo  {journal} {Soft Matter}\ }\textbf {\bibinfo {volume}
  {10}},\ \bibinfo {pages} {446} (\bibinfo {year} {2014})}\BibitemShut
  {NoStop}%
\bibitem [{\citenamefont {Garc{\'\i}a~Daza}\ \emph {et~al.}(2021)\citenamefont
  {Garc{\'\i}a~Daza}, \citenamefont {Puertas}, \citenamefont {Cuetos},\ and\
  \citenamefont {Patti}}]{daza2021microrheology}%
  \BibitemOpen
  \bibfield  {author} {\bibinfo {author} {\bibfnamefont {F.~A.}\ \bibnamefont
  {Garc{\'\i}a~Daza}}, \bibinfo {author} {\bibfnamefont {A.~M.}\ \bibnamefont
  {Puertas}}, \bibinfo {author} {\bibfnamefont {A.}~\bibnamefont {Cuetos}}, \
  and\ \bibinfo {author} {\bibfnamefont {A.}~\bibnamefont {Patti}},\
  }\href@noop {} {\bibfield  {journal} {\bibinfo  {journal} {Journal of Colloid
  and Interface Science}\ } (\bibinfo {year} {2021})}\BibitemShut {NoStop}%
\bibitem [{\citenamefont {Tonti}\ \emph {et~al.}(2021)\citenamefont {Tonti},
  \citenamefont {Garc{\'\i}a~Daza},\ and\ \citenamefont
  {Patti}}]{tonti2021diffusion}%
  \BibitemOpen
  \bibfield  {author} {\bibinfo {author} {\bibfnamefont {L.}~\bibnamefont
  {Tonti}}, \bibinfo {author} {\bibfnamefont {F.~A.}\ \bibnamefont
  {Garc{\'\i}a~Daza}}, \ and\ \bibinfo {author} {\bibfnamefont
  {A.}~\bibnamefont {Patti}},\ }\href@noop {} {\bibfield  {journal} {\bibinfo
  {journal} {Journal of Molecular Liquids}\ }\textbf {\bibinfo {volume}
  {338}},\ \bibinfo {pages} {116640} (\bibinfo {year} {2021})}\BibitemShut
  {NoStop}%
\bibitem [{\citenamefont {Gottschalk}\ \emph {et~al.}(1996)\citenamefont
  {Gottschalk}, \citenamefont {Lin},\ and\ \citenamefont
  {Manocha}}]{gottschalk1996obbtree}%
  \BibitemOpen
  \bibfield  {author} {\bibinfo {author} {\bibfnamefont {S.}~\bibnamefont
  {Gottschalk}}, \bibinfo {author} {\bibfnamefont {M.~C.}\ \bibnamefont {Lin}},
  \ and\ \bibinfo {author} {\bibfnamefont {D.}~\bibnamefont {Manocha}},\
  }\href@noop {} {\bibfield  {journal} {\bibinfo  {journal} {Comp. Graph.}\
  }\textbf {\bibinfo {volume} {30}},\ \bibinfo {pages} {171} (\bibinfo {year}
  {1996})}\BibitemShut {NoStop}%
\bibitem [{\citenamefont {John}\ and\ \citenamefont
  {Escobedo}(2005)}]{john2005phase}%
  \BibitemOpen
  \bibfield  {author} {\bibinfo {author} {\bibfnamefont {B.~S.}\ \bibnamefont
  {John}}\ and\ \bibinfo {author} {\bibfnamefont {F.~A.}\ \bibnamefont
  {Escobedo}},\ }\href@noop {} {\bibfield  {journal} {\bibinfo  {journal} {J.
  Phys. Chem. B}\ }\textbf {\bibinfo {volume} {109}},\ \bibinfo {pages} {23008}
  (\bibinfo {year} {2005})}\BibitemShut {NoStop}%
\bibitem [{\citenamefont {Tonti}\ and\ \citenamefont
  {Patti}(2021)}]{tonti2021fast}%
  \BibitemOpen
  \bibfield  {author} {\bibinfo {author} {\bibfnamefont {L.}~\bibnamefont
  {Tonti}}\ and\ \bibinfo {author} {\bibfnamefont {A.}~\bibnamefont {Patti}},\
  }\href@noop {} {\bibfield  {journal} {\bibinfo  {journal} {Algorithms}\
  }\textbf {\bibinfo {volume} {14}},\ \bibinfo {pages} {72} (\bibinfo {year}
  {2021})}\BibitemShut {NoStop}%
\bibitem [{\citenamefont {Patti}\ and\ \citenamefont
  {Cuetos}(2012)}]{patti2012brownian}%
  \BibitemOpen
  \bibfield  {author} {\bibinfo {author} {\bibfnamefont {A.}~\bibnamefont
  {Patti}}\ and\ \bibinfo {author} {\bibfnamefont {A.}~\bibnamefont {Cuetos}},\
  }\href@noop {} {\bibfield  {journal} {\bibinfo  {journal} {Phys. Rev. E}\
  }\textbf {\bibinfo {volume} {86}},\ \bibinfo {pages} {011403} (\bibinfo
  {year} {2012})}\BibitemShut {NoStop}%
\bibitem [{\citenamefont {Cuetos}\ and\ \citenamefont
  {Patti}(2015)}]{cuetos2015equivalence}%
  \BibitemOpen
  \bibfield  {author} {\bibinfo {author} {\bibfnamefont {A.}~\bibnamefont
  {Cuetos}}\ and\ \bibinfo {author} {\bibfnamefont {A.}~\bibnamefont {Patti}},\
  }\href@noop {} {\bibfield  {journal} {\bibinfo  {journal} {Phys. Rev. E}\
  }\textbf {\bibinfo {volume} {92}},\ \bibinfo {pages} {022302} (\bibinfo
  {year} {2015})}\BibitemShut {NoStop}%
\bibitem [{\citenamefont {Corbett}\ \emph {et~al.}(2018)\citenamefont
  {Corbett}, \citenamefont {Cuetos}, \citenamefont {Dennison},\ and\
  \citenamefont {Patti}}]{corbett2018dynamic}%
  \BibitemOpen
  \bibfield  {author} {\bibinfo {author} {\bibfnamefont {D.}~\bibnamefont
  {Corbett}}, \bibinfo {author} {\bibfnamefont {A.}~\bibnamefont {Cuetos}},
  \bibinfo {author} {\bibfnamefont {M.}~\bibnamefont {Dennison}}, \ and\
  \bibinfo {author} {\bibfnamefont {A.}~\bibnamefont {Patti}},\ }\href@noop {}
  {\bibfield  {journal} {\bibinfo  {journal} {Phys. Chem. Chem. Phys.}\
  }\textbf {\bibinfo {volume} {20}},\ \bibinfo {pages} {15118} (\bibinfo {year}
  {2018})}\BibitemShut {NoStop}%
\bibitem [{\citenamefont {Garc{\'\i}a~Daza}\ \emph {et~al.}(2020)\citenamefont
  {Garc{\'\i}a~Daza}, \citenamefont {Cuetos},\ and\ \citenamefont
  {Patti}}]{daza2020dynamic}%
  \BibitemOpen
  \bibfield  {author} {\bibinfo {author} {\bibfnamefont {F.~A.}\ \bibnamefont
  {Garc{\'\i}a~Daza}}, \bibinfo {author} {\bibfnamefont {A.}~\bibnamefont
  {Cuetos}}, \ and\ \bibinfo {author} {\bibfnamefont {A.}~\bibnamefont
  {Patti}},\ }\href@noop {} {\bibfield  {journal} {\bibinfo  {journal} {Phys.
  Rev. E}\ }\textbf {\bibinfo {volume} {102}},\ \bibinfo {pages} {013302}
  (\bibinfo {year} {2020})}\BibitemShut {NoStop}%
\bibitem [{\citenamefont {Chiappini}\ \emph {et~al.}(2020)\citenamefont
  {Chiappini}, \citenamefont {Patti},\ and\ \citenamefont
  {Dijkstra}}]{chiappini2020}%
  \BibitemOpen
  \bibfield  {author} {\bibinfo {author} {\bibfnamefont {M.}~\bibnamefont
  {Chiappini}}, \bibinfo {author} {\bibfnamefont {A.}~\bibnamefont {Patti}}, \
  and\ \bibinfo {author} {\bibfnamefont {M.}~\bibnamefont {Dijkstra}},\ }\href
  {\doibase 10.1103/PhysRevE.102.040601} {\bibfield  {journal} {\bibinfo
  {journal} {Phys. Rev. E}\ }\textbf {\bibinfo {volume} {102}},\ \bibinfo
  {pages} {040601(R)} (\bibinfo {year} {2020})}\BibitemShut {NoStop}%
\bibitem [{\citenamefont {Carrasco}\ and\ \citenamefont {{Garc{\'\i}a de la
  Torre}}(1999)}]{carrasco1999hydrodynamic}%
  \BibitemOpen
  \bibfield  {author} {\bibinfo {author} {\bibfnamefont {B.}~\bibnamefont
  {Carrasco}}\ and\ \bibinfo {author} {\bibfnamefont {J.}~\bibnamefont
  {{Garc{\'\i}a de la Torre}}},\ }\href@noop {} {\bibfield  {journal} {\bibinfo
   {journal} {Biophys. J.}\ }\textbf {\bibinfo {volume} {76}},\ \bibinfo
  {pages} {3044} (\bibinfo {year} {1999})}\BibitemShut {NoStop}%
\bibitem [{\citenamefont {{Garc{\'\i}a de la Torre}}\ \emph
  {et~al.}(2007)\citenamefont {{Garc{\'\i}a de la Torre}}, \citenamefont {del
  Rio~Echenique},\ and\ \citenamefont {Ortega}}]{garcia2007improved}%
  \BibitemOpen
  \bibfield  {author} {\bibinfo {author} {\bibfnamefont {J.}~\bibnamefont
  {{Garc{\'\i}a de la Torre}}}, \bibinfo {author} {\bibfnamefont
  {G.}~\bibnamefont {del Rio~Echenique}}, \ and\ \bibinfo {author}
  {\bibfnamefont {A.}~\bibnamefont {Ortega}},\ }\href@noop {} {\bibfield
  {journal} {\bibinfo  {journal} {J. Phys. Chem. B}\ }\textbf {\bibinfo
  {volume} {111}},\ \bibinfo {pages} {955} (\bibinfo {year}
  {2007})}\BibitemShut {NoStop}%
\bibitem [{\citenamefont {Squires}\ and\ \citenamefont
  {Brady}(2005)}]{squires2005simple}%
  \BibitemOpen
  \bibfield  {author} {\bibinfo {author} {\bibfnamefont {T.~M.}\ \bibnamefont
  {Squires}}\ and\ \bibinfo {author} {\bibfnamefont {J.~F.}\ \bibnamefont
  {Brady}},\ }\href@noop {} {\bibfield  {journal} {\bibinfo  {journal} {Physics
  of Fluids}\ }\textbf {\bibinfo {volume} {17}},\ \bibinfo {pages} {073101}
  (\bibinfo {year} {2005})}\BibitemShut {NoStop}%
\bibitem [{\citenamefont {Khair}\ and\ \citenamefont
  {Brady}(2006)}]{khair2006single}%
  \BibitemOpen
  \bibfield  {author} {\bibinfo {author} {\bibfnamefont {A.~S.}\ \bibnamefont
  {Khair}}\ and\ \bibinfo {author} {\bibfnamefont {J.~F.}\ \bibnamefont
  {Brady}},\ }\href@noop {} {\bibfield  {journal} {\bibinfo  {journal} {Journal
  of Fluid Mechanics}\ }\textbf {\bibinfo {volume} {557}},\ \bibinfo {pages}
  {73} (\bibinfo {year} {2006})}\BibitemShut {NoStop}%
\bibitem [{\citenamefont {Gazuz}\ \emph {et~al.}(2009)\citenamefont {Gazuz},
  \citenamefont {Puertas}, \citenamefont {Voigtmann},\ and\ \citenamefont
  {Fuchs}}]{gazuz2009active}%
  \BibitemOpen
  \bibfield  {author} {\bibinfo {author} {\bibfnamefont {I.}~\bibnamefont
  {Gazuz}}, \bibinfo {author} {\bibfnamefont {A.~M.}\ \bibnamefont {Puertas}},
  \bibinfo {author} {\bibfnamefont {T.}~\bibnamefont {Voigtmann}}, \ and\
  \bibinfo {author} {\bibfnamefont {M.}~\bibnamefont {Fuchs}},\ }\href@noop {}
  {\bibfield  {journal} {\bibinfo  {journal} {Physical review letters}\
  }\textbf {\bibinfo {volume} {102}},\ \bibinfo {pages} {248302} (\bibinfo
  {year} {2009})}\BibitemShut {NoStop}%
\bibitem [{\citenamefont {Swan}\ and\ \citenamefont
  {Zia}(2013)}]{swan2013active}%
  \BibitemOpen
  \bibfield  {author} {\bibinfo {author} {\bibfnamefont {J.~W.}\ \bibnamefont
  {Swan}}\ and\ \bibinfo {author} {\bibfnamefont {R.~N.}\ \bibnamefont {Zia}},\
  }\href@noop {} {\bibfield  {journal} {\bibinfo  {journal} {Physics of
  Fluids}\ }\textbf {\bibinfo {volume} {25}},\ \bibinfo {pages} {083303}
  (\bibinfo {year} {2013})}\BibitemShut {NoStop}%
\bibitem [{\citenamefont {Puertas}\ and\ \citenamefont
  {Voigtmann}(2014)}]{puertas2014microrheology}%
  \BibitemOpen
  \bibfield  {author} {\bibinfo {author} {\bibfnamefont {A.~M.}\ \bibnamefont
  {Puertas}}\ and\ \bibinfo {author} {\bibfnamefont {T.}~\bibnamefont
  {Voigtmann}},\ }\href@noop {} {\bibfield  {journal} {\bibinfo  {journal}
  {Journal of Physics: Condensed Matter}\ }\textbf {\bibinfo {volume} {26}},\
  \bibinfo {pages} {243101} (\bibinfo {year} {2014})}\BibitemShut {NoStop}%
\bibitem [{\citenamefont {Meyer}\ \emph {et~al.}(2006)\citenamefont {Meyer},
  \citenamefont {Marshall}, \citenamefont {Bush},\ and\ \citenamefont
  {Furst}}]{meyer2006laser}%
  \BibitemOpen
  \bibfield  {author} {\bibinfo {author} {\bibfnamefont {A.}~\bibnamefont
  {Meyer}}, \bibinfo {author} {\bibfnamefont {A.}~\bibnamefont {Marshall}},
  \bibinfo {author} {\bibfnamefont {B.~G.}\ \bibnamefont {Bush}}, \ and\
  \bibinfo {author} {\bibfnamefont {E.~M.}\ \bibnamefont {Furst}},\ }\href@noop
  {} {\bibfield  {journal} {\bibinfo  {journal} {Journal of rheology}\ }\textbf
  {\bibinfo {volume} {50}},\ \bibinfo {pages} {77} (\bibinfo {year}
  {2006})}\BibitemShut {NoStop}%
\bibitem [{\citenamefont {Sriram}\ \emph {et~al.}(2010)\citenamefont {Sriram},
  \citenamefont {Meyer},\ and\ \citenamefont {Furst}}]{sriram2010active}%
  \BibitemOpen
  \bibfield  {author} {\bibinfo {author} {\bibfnamefont {I.}~\bibnamefont
  {Sriram}}, \bibinfo {author} {\bibfnamefont {A.}~\bibnamefont {Meyer}}, \
  and\ \bibinfo {author} {\bibfnamefont {E.~M.}\ \bibnamefont {Furst}},\
  }\href@noop {} {\bibfield  {journal} {\bibinfo  {journal} {Physics of
  Fluids}\ }\textbf {\bibinfo {volume} {22}},\ \bibinfo {pages} {062003}
  (\bibinfo {year} {2010})}\BibitemShut {NoStop}%
\bibitem [{\citenamefont {Paladugu}\ \emph {et~al.}(2020)\citenamefont
  {Paladugu}, \citenamefont {Kaur}, \citenamefont {Mohiuddin}, \citenamefont
  {Pujala}, \citenamefont {Pal},\ and\ \citenamefont
  {Dhara}}]{paladugu2020microrheology}%
  \BibitemOpen
  \bibfield  {author} {\bibinfo {author} {\bibfnamefont {S.}~\bibnamefont
  {Paladugu}}, \bibinfo {author} {\bibfnamefont {S.}~\bibnamefont {Kaur}},
  \bibinfo {author} {\bibfnamefont {G.}~\bibnamefont {Mohiuddin}}, \bibinfo
  {author} {\bibfnamefont {R.~K.}\ \bibnamefont {Pujala}}, \bibinfo {author}
  {\bibfnamefont {S.~K.}\ \bibnamefont {Pal}}, \ and\ \bibinfo {author}
  {\bibfnamefont {S.}~\bibnamefont {Dhara}},\ }\href@noop {} {\bibfield
  {journal} {\bibinfo  {journal} {Soft Matter}\ }\textbf {\bibinfo {volume}
  {16}},\ \bibinfo {pages} {7556} (\bibinfo {year} {2020})}\BibitemShut
  {NoStop}%
\end{thebibliography}%

\end{document}